\documentclass[twocolumn]{aastex6}

\pdfoutput=1

\usepackage{amsmath,amssymb,amstext}
\usepackage{aas_macros} % 
\usepackage{hyperref}  % creates hyperlinks within doc
\usepackage{graphicx} % imports graphics
\usepackage{bm}

\shorttitle{GMC Collisions as Triggers of Star Formation. III.}
\shortauthors{Wu et al.}

\begin{document}

\title{GMC Collisions as Triggers of Star Formation. III. \\
Density and Magnetically Regulated Star Formation}

\author{Benjamin Wu}
\affil{National Astronomical Observatory of Japan, Mitaka, Tokyo 181-8588, Japan}
\affil{Department of Physics, University of Florida, Gainesville, FL 32611, USA}
\email{ben.wu@nao.ac.jp}
\author{Jonathan C. Tan}
\affil{Department of Physics, University of Florida, Gainesville, FL 32611, USA}
\affil{Department of Astronomy, University of Florida, Gainesville, FL 32611, USA}
\author{Duncan Christie}
\affil{Department of Astronomy, University of Florida, Gainesville, FL 32611, USA}
\author{Fumitaka Nakamura}
\affil{National Astronomical Observatory, Mitaka, Tokyo 181-8588, Japan}
\author{Sven Van Loo}
\affil{School of Physics and Astronomy, University of Leeds, Leeds LS2 9JT, UK}
\and
\author{David Collins}
\affil{Department of Physics, Florida State University, Tallahassee, FL 32306-4350, USA}

\begin{abstract}
We study giant molecular cloud (GMC) collisions and their ability to
trigger star cluster formation. We further develop our three
dimensional magnetized, turbulent, colliding GMC simulations by
implementing star formation sub-grid models. Two such models are
explored: (1) ``Density-Regulated,'' i.e., fixed efficiency per
free-fall time above a set density threshold; (2) ``Magnetically-
Regulated,'' i.e., fixed efficiency per free-fall time in regions that
are magnetically supercritical. Variations of parameters associated
with these models are also explored. In the non-colliding simulations,
the overall level of star formation is sensitive to model parameter
choices that relate to effective density thresholds. In the GMC
collision simulations, the final star formation rates and efficiencies
are relatively independent of these parameters. Between non-colliding
and colliding cases, we compare the morphologies of the resulting star
clusters, properties of star-forming gas, time evolution of the star
formation rate (SFR), spatial clustering of the stars, and resulting
kinematics of the stars in comparison to the natal gas. We find that
typical collisions, by creating larger amounts of dense gas, trigger
earlier and enhanced star formation, resulting in 10 times higher SFRs
and efficiencies. The star clusters formed from GMC collisions show
greater spatial sub-structure and more disturbed kinematics.
\end{abstract}


\section{Introduction}
\label{sec:intro}

Most stars are thought to form in clusters within giant molecular
clouds (GMCs). GMCs have typical hydrogen number densities of $n_{\rm
  H}=100\:{\rm cm^{-3}}$, diameters of $\sim$tens of parsecs, masses
of up to $10^{6}\:M_{\odot}$, and average temperatures of $\sim
10-30\:{\rm K}$. Dense clumps within GMCs, potentially traced as,
e.g., Infrared Dark Clouds (IRDCs), are recognized as being the likely
precursors to star clusters
\citep[e.g.][]{Rathborne_ea_2006,Tan_ea_2014,Butler_Tan_2009,Butler_Tan_2012}. 
IRDCs have such high mass surface densities ($\Sigma \gtrsim 0.1~{\rm
  g\:cm^{-2}}$) that they are dark at mid-IR ($\sim10~{\rm \mu m}$)
and even far-IR ($\sim70~{\rm \mu m}$) \citep[e.g., ][]{Lim_Tan_2014}.  
Their low temperatures ($10-20$K) \citep[see,
  e.g.,][]{Pillai_ea_2006,Wang_ea_2008,Sakai_ea_2008,Chira_ea_2013},
high volume densities ($n_{\rm H}>10^{5}~{\rm cm^{-3}}$), relatively
compact sizes ($\sim$few pc), and masses ($\sim
10^{2}-10^{5}~M_{\odot}$) indicate that they have the potential to be
the precursors to most of the observed mass range of star clusters
known in the Galaxy. The initial and early stages of star cluster
formation can also be traced by dust continuum emission 
\citep[e.g.,][]{Ginsburg_ea_2012} and by samples based on emission of dense gas
tracers \citep[e.g.,][]{Ma_ea_2013}. Surveys of young embedded stars also
can probe the structure \citep[e.g.,][]{Jaehnig_ea_2015}, age distribution
\citep[e.g.,][]{DaRio_ea_2014} and kinematics of young clusters 
\citep[e.g.,][]{Foster_ea_2015,Cottaar_ea_2015}.

Currently, the dominant processes that induce the collapse and
fragmentation of GMCs into star-forming clumps are
poorly understood. 
Various theoretical models include 
regulation by turbulence \citep[e.g.,][]{Krumholz_McKee_2005},
regulation by magnetic fields \citep[e.g.,][]{VanLoo_ea_2015},
triggering by stellar feedback \citep[e.g.,
  supernova;][]{Inutsuka_ea_2015}, triggering by converging atomic
flows \citep[e.g.,][]{Heitsch_Stone_Hartmann_2009}, and triggering via
converging molecular flows, i.e., GMC-GMC collisions
\citep[e.g.,][]{Scoville_Sanders_Clemens_1986,Tan_2000}.

Semi-analytic models \citep{Tan_2000} and numerical simulations
\citep{Tasker_Tan_2009,Fujimoto_ea_2014,Dobbs_ea_2015} of global
galactic disks have shown that GMCs collide relatively frequently due
to the approximately 2D geometry of a thin disk and interaction rates
driven by differential rotation of galactic orbits. The average
timescale between GMC collisions was found to be about 20\% of a local
orbital period within a flat rotation curve disk
\citep{Tasker_Tan_2009} \citep[see also][]{Fujimoto_ea_2014,Dobbs_ea_2015}.
A growing number of numerical studies have also
shown that collisions between molecular clouds can provide conditions
favorable for massive star and star cluster formation \citep[see
  e.g.,][]{Habe_Ohta_1992,Klein_Woods_1998,Anathpindika_2009,Takahira_ea_2014,Haworth_ea_2015a,Haworth_ea_2015b,Balfour_ea_2015}.
We note that in general comparison of the results between the
  simulations of different groups is complicated by the use of
  different initial conditions, different numerical methods and
  different included physics.

Our approach here is to systematically build-up realism for our
GMC collision simulations by including additional physics step by
step that allows an understanding of the relative importance of
different input assumptions. \citet[][hereafter Paper I]{Wu_ea_2015} 
and \citet[][hereafter Paper II]{Wu_ea_2017}
developed a numerical study of GMC-GMC collisions, focusing on
understanding the physical mechanisms as well as using them to predict
observational diagnostics. Comparing magnetized, supersonically
turbulent GMCs in colliding and non-colliding cases over a wide
parameter space and investigating a varied array of potential
observational signatures, they found that a number of indicators
suggest similarities between the colliding scenarios to observed GMCs
and IRDCs. Further, dynamical virial analysis suggested that dense
$^{13}$CO-defined structures created through GMC collisions were more
likely to collapse and form massive star clusters when compared with
more quiescently evolving structures.

The next stage in our work is the crucial transition from collapsing
clumps into star clusters. Properties of the stars that form, along
with their dynamical evolution shortly thereafter, may provide insight
into the dominant star formation mechanisms.  The goal of this study
is to answer the question: do realistic models of GMC collisions
create star clusters that closely match the properties of observed
young star-forming regions?  We approach this question by further
building upon our previous numerical framework of GMCs through the
development of star formation sub-grid models, one of which is a novel
magnetically-regulated model. We combine our existing gas-focused
observational diagnostic methods with additional information from the
population of star particles. Thus, we hope to provide insight to the
star formation process by analyzing the evolution of IRDC-type
structures into young star clusters.

Section \ref{sec:methods} describes our numerical setup and the
various star formation models.  We then present our results in
Section \ref{sec:results}, which include: 
gas and star cluster morphologies (\S\ref{sec:results-morph}), 
properties of star-forming gas (\S\ref{sec:results-SFcells}),
global star formation rates (SFRs) and efficiencies 
(\S\ref{sec:results-SFR}), spatial clustering (\S\ref{sec:results-clust}), 
and star particle kinematics (\S\ref{sec:results-SFppv}). 
In Section \ref{sec:conclusion} we discuss our conclusions.

\section{Numerical Model}
\label{sec:methods}

\subsection{Initial Conditions}

We further develop the numerical framework described in Paper II and
introduce two star formation routines.  Our GMCs are identical to
those initialized in Paper II, which are motivated by observed GMC
properties. The clouds are self-gravitating, supersonically turbulent,
and magnetized. They are initialized with identical densities and
offset by an impact parameter. The clouds are embedded in an ambient
medium of ten times lower density (i.e., an atomic cold neutral
medium, CNM), which for the colliding case, is converging along with
the GMCs. The initial simulation properties are summarized in
Table~\ref{tab:initial}.

The simulation domain is $(128\:{\rm pc})^{3}$ and contains two
neighboring GMCs. The GMCs are initially uniform spheres, with
Hydrogen number densities of $n_{\rm H,GMC} = 100\:{\rm cm^{-3}}$ and
radii $R_{\rm GMC} = 20.0$~pc. This gives each GMC a mass $M_{\rm GMC}
= 9.3 \times 10^4\:M_\odot$.  The ambient gas represents the atomic
cold neutral medium (CNM) and has a density of $n_{\rm H,0} = 10\:{\rm
  cm^{-3}}$. The centers of the GMCs are offset by $2R_{\rm GMC}$ in
the collision axis ($x$), 0 in the $y$-axis, and $b=0.5R_{\rm GMC}$ in
the $z$-axis.

To approximate the density and velocity structures observed in GMCs,
our clouds are initialized with a supersonic turbulent velocity field
which is random, purely solenoidal, and follows the $v_{k}^{2} \propto
k^{-4}$ relation, where $k=\pi/d$ is the wavenumber for an eddy
diameter $d$. Conventionally, the ``$k$-mode'' is normalized to the
simulation box length. The gas within the GMC is initialized with Mach
number $\mathcal{M}_{s} \equiv \sigma/ c_{s}= 23$ (for $T=15$~K
conditions), of order virial. We set our fiducial $k$-modes to be
$\left\{2,...,20\right\}$, where each mode within this range is
excited.  This is representative of the large-scale turbulent
velocities (small $k$) spanning from the GMC diameters down to a small
enough minimum scale (large $k$), which is numerically resolved, but
expected to cascade to smaller scales. We do not drive turbulence,
instead letting it decay within a few dynamical times.  Note also that
turbulence is initialized only within the initial volume of the GMCs
while we leave the ambient medium non-turbulent. Note also the GMC
collision will also drive turbulence in the clouds in that case.

A large-scale uniform magnetic field of strength $10\:{\rm \mu G}$ is
initialized throughout the box at an angle $\theta=60^{\circ}$ with
respect to the collision ($x$-) axis. This choice of $|\bm{B}|$ is
motivated by the Zeeman measurements of typical GMC field strengths,
summarized by \citet{Crutcher_2012}.

In the fiducial colliding case, the bulk flows (including both the
ambient gas and the GMCs) have a relative velocity of $v_{\rm
  rel}=10\:{\rm km\:s^{-1}}$. In the non-colliding case, there is no
bulk velocity flow.

The simulations are run for $5\:{\rm Myr}$ to investigate the onset of
star formation. Note that this is $1\:{\rm Myr}$ longer than the
simulations described in Paper II, which focused on gas properties of
the pre-star-forming clump. Note also that the freefall time given the
initial uniform density GMCs is $t_{\rm
  ff}=(3\pi/[32G\rho])^{1/2}\simeq 4.35$~Myr. However, the values of
$t_{\rm ff}$ for the denser substructures created by turbulence and by
the collision are much shorter.
Star formation is expected to occur in both non-colliding and 
colliding cases, with the detailed properties of resulting star 
clusters acting as the key point of our investigation.

\begin{deluxetable}{llllclc}
\tablecaption{Initial Simulation Properties \label{tab:initial}}
\tablecolumns{4}
\tablewidth{0pt}
\tablehead{
\colhead{ } & \colhead{ } & \colhead{ } & \colhead{ } &
\colhead{GMC} & \colhead{ } &\colhead{Ambient} \\
}
\startdata
$n_{\rm H}$    & & (${\rm cm}^{-3}$) & & 100               & & 10      \\
$R$            & & (${\rm pc}$)      & & 20                & &  ...    \\
$M$            & & ($M_{\odot}$)     & & $9.3\times10^{4}$ & &  ...    \\
$T$            & & (K)               & & 15                & & 150     \\
$t_{\rm ff}$   & & (Myr)             & & 4.35              & &  ...    \\
$c_{\rm s}$    & & (km/s)            & & 0.23              & & 0.72    \\
$v_{A}$        & & (km/s)            & & 1.84              & & 5.83    \\
$v_{\rm vir}$  & & (km/s)            & & 4.9               & &  ...    \\
$\sigma$       & & (km/s)            & & 5.2               & &  ...    \\
$\mathcal{M}_{\rm s}$ & &    ...     & & 23                & &  ...    \\ 
$\mathcal{M}_{A}$ & &       ...      & & 2.82              & &  ...    \\ 
$k$-mode       & & ($k_{1},k_{2}$)   & & (2,20)            & &  ...    \\
$v_{\rm bulk}$ & & (km/s)            & & $\pm5$            & & $\pm5$  \\
$B$            & & (${\rm \mu G}$)   & & 10                & & 10      \\
$\lambda$\tablenotemark{a} & &  ...  & & 4.3               & & 1.5     \\
$\beta$\tablenotemark{b}   & &  ...  & & 0.015             & & 0.015   \\
\enddata
\tablenotetext{a}{normalized mass-to-flux ratio: $\lambda = (M/\Phi)/(1/2\pi G^{1/2})$}
\tablenotetext{b}{thermal-to-magnetic pressure ratio: $\beta=8\pi c_{s}^2 \rho_{0} / B^2$}
\end{deluxetable}

\begin{figure}
\centering
\includegraphics[width=1\columnwidth]{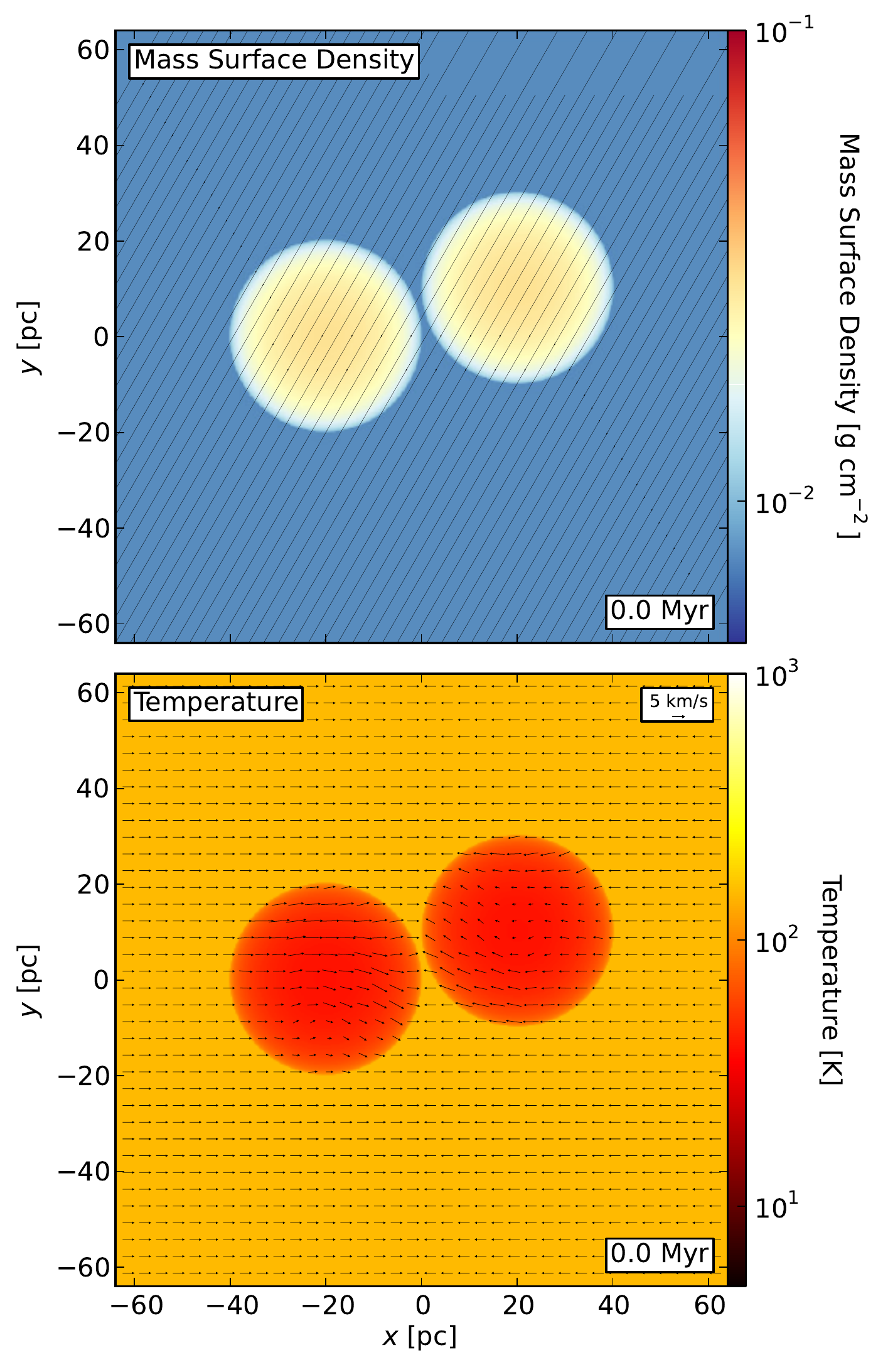}
\caption{
Initial conditions. Top panel: Mass surface density, shown together
with magnetic field structure (gray lines).  Bottom panel:
Mass-weighted temperature, shown together with the velocity field
(black vectors). The colliding case is shown.  GMCs 1 (left) and 2
(right) have identical dimensions with an initial separation of their
centers of $2R_{\rm GMC}$ in the x-direction and 0 in the
z-direction. In the y-direction, they are offset by an impact
parameter $b=0.5R_{\rm GMC}$.
\label{fig:ICs}}
\end{figure}

\subsection{Numerical Code}
\label{sec:methods-code}

Our models are run using
\texttt{Enzo}\footnote{http://enzo-project.org (v2.4)}, a
magnetohydrodynamics (MHD) adaptive mesh refinement (AMR) code
\citep{Bryan_ea_2014}. We use the Dedner-MHD method, which solves the
solves the MHD equations using the Harten-Lax-van Leer with
Discontinuities (HLLD) method and a piecewise linear reconstruction
method (PLM). The time is evolved using the MUSCL 2nd-order
Runge-Kutta method. The $\nabla \cdot \bf{B}=0$ solenoidal constraint
of the magnetic field is maintained via a hyperbolic divergence
cleaning method \citep{Dedner_ea_2002,Wang_Abel_2008}.

The simulation domain is realized with a top level root grid of
$128^{3}$ with 3 additional levels of AMR. Our models thus have an
effective resolution of $1024^{3}$, with a minimum grid cell size of
0.125~pc. We refine solely on the local Jeans length, setting a
necessary requirement of resolving by 8 cells, i.e., higher than the 4
cells typically used to avoid artificial fragmentation
\citep{Truelove_ea_1997}). Our higher resolution leads to larger
volumes of the GMCs being better resolved and thus generally better
resolution of, e.g., shocks \citep[see also][]{Few_ea_2016}.
We note that the Jeans criterion assumes purely thermal support. The
gas in our simulations also has some magnetic support, so its
effective ``magneto-Jeans length'' will be significantly larger than
the thermal Jeans length in directions perpendicular to the magnetic
field.

Due to the relatively high bulk velocities and 
potentially strong magnetic fields, we require the use of the
{\textquotedblleft}dual energy formalism{\textquotedblright}
\citep{Bryan_ea_2014} , which separately solves the internal
energy equation as well as the total energy equation, ensuring
  accurate calculation of pressures and temperatures in these
  conditions. If the ratio of thermal to total energy is less than
0.001, then the temperature is calculated from the internal
pressure. Otherwise, the total energy is used.

Additionally, we employ the ``Alfv\'en limiter'' (described in Paper
II) to avoid exceedingly small timesteps set by Alfv\'en waves. This
acts by choosing a maximum Alfv\'en velocity, $v_{A,{\rm
    max}}=B/\sqrt{4\pi\rho_{\rm min}}=1\times10^{7}\:{\rm cm
  \:s^{-1}}$, and setting a density floor that is determined by the
magnetic field. This predominantly affects only small pockets of very
low-density gas with which we are less interested, and thus the
dynamical results are deemed unaffected by this limiter.

\subsection{Thermal Processes}
\label{sec:methods-thermal}

We assume a constant mean particle mass ($\mu=2.33\:m_{\rm H}$)
throughout the simulation domain for simplicity, as our focus is on
the dense molecular gas of GMCs.  We also choose a constant adiabatic
index $\gamma=5/3$. Note that this essentially ignores certain
excitation modes of $\rm H_{2}$ that may be relevant (i.e., shocks),
but it is still the most appropriate single-valued choice of $\gamma$,
given our focus on the dynamics of cold $\rm H_{2}$. Also, we assume
$n_{\rm He}=0.1n_{\rm H}$, giving a mass per H of
$2.34\times10^{-24}\:{\rm g}$.

The PDR-based heating and cooling functions developed in Paper I are
again used in these simulations. The assumptions are: (1) FUV
radiation field of $G_{0}=4$ (i.e., appropriate conditions for the
inner Galaxy, e.g., at Galactocentric distances of $\sim 4$~kpc) and
(2) a background cosmic ray ionization rate of $\zeta = 1.0 \times
10^{-16}\:{\rm s^{-1}}$.  The heating/cooling functions trace the
atomic to molecular transition and recreate a multi-phase ISM. They
span density and temperature ranges of $10^{-3} \leq n_{\rm H}/{\rm
  cm^{-3}} \leq 10^{6}$ (extended to $10^10~{\rm cm^{-3}}$ via
extrapolation) and $2.7 \leq T/{\rm K} \leq 10^7$, respectively.

We use the \texttt{Grackle} external chemistry and cooling
library\footnote{https://grackle.readthedocs.org/}
\citep{Smith_ea_2016arXiv} to incorporate our heating/cooling
functions in tabular form into \texttt{Enzo}, modifying the energy
equation.

In order to avoid numerical instabilities related to the
heating/cooling processes, we limit the timestep on each AMR level to
a factor of 0.2 the minimum cooling time. Additionally, we set a hard
floor for the minimum cooling timestep of $625~{\rm yrs}$.

\begin{table*}
  \caption{Simulation Runs with Star Formation Sub-Grid Models}
  \label{tab:SFmodels}
  \centering
  \begin{tabular}{lcccccccc} \tableline \tableline
Name & Star Formation & $v_{\rm rel}$ &  Cell Size &  $n_{\rm H,sf}$  & $t_{\rm ff}$ & $m_{\rm gas,min}$  &  $m_{\rm \star, min}$  & $c_{1}$  \\
     & Model          & (${\rm km\:s^{-1}}$) & (pc)     &  (cm$^{-3}$)     & (yr)         &  ($M_{\odot}$)    &  ($M_{\odot}$)   &       \\
\tableline
d-0.5-nocol & dens. reg. & 0 & 0.125 & $0.5\times 10^{6}$ & $6.2\times 10^{4}$ & 32 & 10 & - \\
d-1-nocol & dens. reg. & 0 & 0.125 & $1.0\times 10^{6}$ & $4.4\times 10^{4}$ & 63 & 10 & - \\
d-2-nocol & dens. reg. & 0 & 0.125 & $2.0\times 10^{6}$ & $3.1\times 10^{4}$ & 126 & 10 & - \\
B-0.5-nocol & mag. reg. & 0 & 0.125 &  $3.55\times10^{5}$  & $7.3\times 10^{4}$  & 20  & 10 & 0.063     \\
B-1-nocol & mag. reg. & 0 & 0.125 &  $3.55\times10^{5}$  & $7.3\times 10^{4}$  & 20  & 10 & 0.126     \\
B-2-nocol & mag. reg. & 0 & 0.125 &  $3.55\times10^{5}$  & $7.3\times 10^{4}$  & 20  & 10 & 0.252     \\
B-1-1M-nocol & mag. reg. & 0 & 0.125 &  $3.55\times10^{4}$  & $2.3\times 10^{5}$  & 2  & 1 & 0.126     \\
\tableline
d-0.5-col & dens. reg. & 10 & 0.125 & $0.5\times 10^{6}$ & $6.2\times 10^{4}$ & 32 & 10 & -  \\
d-1-col & dens. reg. & 10 & 0.125 & $1.0\times 10^{6}$ & $4.4\times 10^{4}$ & 63 & 10 & -  \\
d-2-col & dens. reg. & 10 & 0.125 & $2.0\times 10^{6}$ & $3.1\times 10^{4}$ & 126 & 10 & -  \\
B-0.5-col & mag. reg. & 10 & 0.125 &  $3.55\times10^{5}$  & $7.3\times 10^{4}$  & 20  & 10 & 0.063     \\
B-1-col & mag. reg. & 10 & 0.125 &  $3.55\times10^{5}$  & $7.3\times 10^{4}$  & 20  & 10 & 0.126     \\
B-2-col & mag. reg. & 10 & 0.125 &  $3.55\times10^{5}$  & $7.3\times 10^{4}$  & 20  & 10 & 0.252     \\
B-1-1M-col & mag. reg. & 10 & 0.125 &  $3.55\times10^{4}$  & $2.3\times 10^{5}$  & 2  & 1 & 0.126     \\
\tableline
\end{tabular}
\end{table*}

\subsection{Star Formation}
\label{sec:methods-SF}

We utilize the particle machinery of \texttt{Enzo} to model star
formation.  Specifically, star particles (i.e., collisionless, point
particles with mass $m_{\star}$) form within a simulation cell if
certain local criteria are met.  Two star formation routines are
developed: (1) density-regulated star formation; (2)
magnetically-regulated star formation.

\subsubsection{Density-regulated star formation}\label{S:denSF}

Our first star formation routine is a ``density-regulated'' model,
based on that of \citet{Van_Loo_Butler_Tan_2013} \citep[see also][]{Butler_ea_2015}.
Stars are formed within a cell only if they have been
refined to the finest level of resolution and the density exceeds a
particular threshold value, $n_{\rm H,sf}$. The fiducial star
formation density threshold is chosen to be $n_{\rm H,sf} =
10^{6}~{\rm cm^{-3}}$, which is set partly based on observed densities
of pre-stellar cores \citep[e.g.,][]{Kong_ea_2017arXiv}.
We will consider variation of this parameter by a factor of two to
higher and lower values. The temperature in the cell is also required
to be $<3000$~K, to avoid star formation in dense, shock-heated
regions, but we will see that this constraint is not of practical
concern for the simulations presented here. Note that there is no
requirement for gravitational boundedness of gas in the cell. Nor is
there a requirement for net convergence of gas flow to the cell. These
choices are motivated by the fact such conditions are not well
resolved on the local scales associated with an individual cell. In
addition, we expect that processes such as turbulence and diffusion of
magnetic flux that are occurring on sub-grid scales (or scales near the
grid scale that are not well resolved) will regulate star formation,
e.g., creating local conditions that are gravitationally unstable,
perhaps via converging flows. With these points in mind, this star
formation sub-grid model using a density threshold is thus designed to
be as simple as possible, enabling us to gain a clear understanding of
how the results depend on its input parameters.

In cells meeting the above conditions, star particles are then
produced so that the SFR is, on average, equal to that expected if
there is a fixed star formation efficiency per local free-fall time,
$\epsilon_{\rm ff}$, where the local free-fall time, $t_{\rm ff}$, is
expressed as 
\begin{eqnarray}
t_{\rm ff} & = & \left(\frac{3\pi}{32G\rho}\right)^{1/2} \\ & = & 4.4\times10^{4} n_{\rm H,6}^{-1/2}\:{\rm yr},
\end{eqnarray}
i.e., the value for collapse of a uniform density sphere, where
$n_{\rm H,6}\equiv n_{\rm H}/10^6\:{\rm cm^{-3}}$. We adopt a fiducial
choice of $\epsilon_{\rm ff}=0.02$, motivated by observations of GMCs,
their star-forming clumps and stellar populations in embedded
clusters, which suggest fairly low and density-independent values of
$\epsilon_{\rm ff}$ \citep[see,
  e.g.,][]{Zuckerman_Evans_1974,Krumholz_Tan_2007,DaRio_ea_2014}. 
Thus the SFR is
\begin{eqnarray}
\dot{m}_\star & = & \epsilon_{\rm ff}\frac{m_{\rm gas}}{t_{\rm ff}} \\ & = & 2.9\times10^{-5} \left(\frac{\epsilon_{\rm ff}}{0.02}\right) \left(\frac{\Delta x}{0.125\:{\rm pc}}\right)^3 n_{\rm H,6}^{3/2}\:M_\odot \:{\rm yr}^{-1},
\end{eqnarray}
where we have normalized to the minimum cell size, $\Delta x$,
relevant for the simulations in this paper.

The timesteps in the simulation are typically quite short, i.e., much
less than the signal crossing time of a cell, i.e., $\ll
1.2\times10^5\:$yr for a signal speed of 1~$\rm km\:s^{-1}$. Thus the
average mass of stars that are expected to be created in a given cell
in a given simulation timestep is often very small, i.e.,
$<1\:M_\odot$.

To enable both a practical computation that does not involve too many
star particles, but also with the eventual aim of producing star
particles with masses that are characteristic of observed stellar
masses, the star formation sub-grid model also involves a parameter of
a minimum star particle mass, $m_{\rm \star,min}$. For the
density-regulated models we consider here, we set $m_{\rm
  \star,min}=10\:M_\odot$. Thus in this case the star particles
represent small (sub-)clusters of stars, since the mean stellar mass
is $\lesssim 1\:M_\odot$ for realistic stellar initial mass functions
\citep[e.g.,][]{Parravano_ea_2011}
(see also discussion of star particle
dynamics in \S\ref{S:stardyn}). With this value of $m_{\rm \star,min}$
we are almost always in a regime in which the mass of stars to be
created in a given timestep is smaller than $m_{\rm \star,min}$ and so
the decision to form a star particle or not needs to be implemented
probabilistically, i.e., the ``stochastic star formation'' regime. In
this case, the star particle is formed with probability $\dot{m}_\star
\Delta t / m_{\rm \star,min}$, where $\Delta t$ is the simulation
timestep. If on the other hand $\dot{m}_\star \Delta t > m_{\rm
  \star,min}$ (which can occur in certain circumstances), then the
star particle is simply created with this mass.

Another factor affecting the choice of $m_{\rm \star,min}$ is the
desire not to change the gas mass in a cell by too large a fraction
when the star particle is created, i.e., to avoid too large changes in
density, pressure, etc. In general we set an upper limit of this
fraction of 0.5. In the fiducial case a cell of size 0.125~pc at the
star formation threshold density contains a minimum gas mass of
$m_{\rm gas,min}=63\:M_\odot$, so this fraction is $\lesssim 0.17$ for
these models ($\lesssim 0.34$ for the lower threshold density case).

Overall there are three density (``d'')-regulated runs (i.e., three
choices of threshold density) for each of the noncolliding (``nocol'')
and colliding (``col'') simulation set-ups. The parameters of these
star formation models and simulations are listed in
Table~\ref{tab:SFmodels}.

\subsubsection{Magnetically-regulated star formation}\label{S:magSF}

We introduce a new ``magnetically-regulated'' star formation model
that takes into account magnetic criticality, i.e., star formation is
only allowed to proceed if a cell has a mass to flux ratio that is
greater than a certain value. If the cell is magnetically
``supercritical'' by this criterion, then it forms stars at a fixed
efficiency per local free-fall time, $\epsilon_{\rm ff}$, where we
will adopt the same value of 0.02 that was used in the
density-regulated models. Thus this magnetic criticality condition
acts to replace the density-threshold criterion of
\S\ref{S:denSF}. However, as we discuss below, the choice of $m_{\rm
  \star,min}$ also introduces an effective minimum density for star
formation in this model also.

To assess the mass-to-flux ratio criterion, as an approximation, we
treat each grid cell individually and calculate the dimensionless
mass-to-flux ratio
\begin{equation}
\mu_{\rm cell} = \frac{\rho \Delta x \sqrt{G}}{B c_{1}},
\end{equation}
where $\rho$ is the density within a cell of length $\Delta x$, $G$ is
the gravitational constant, $B$ is the strength of the magnetic field
within the cell, and $c_{1}$ comes from defining the critical
mass-to-flux ratio as
\begin{equation}
\left( \frac{M}{\Phi} \right)_{\rm crit} = \frac{c_{1}}{\sqrt{G}}
\end{equation}
and is ultimately dependent of the geometry of the system. For an
infinite disk, the value is $c_{1}=1/(2\pi)$
\citep{Nakano_Nakamura_1978}; for an isolated cloud it is roughly
$1/\sqrt{63}\sim 0.126$ \citep{Mouschovias_Spitzer_1976}. We will
consider variations of $c_1$ of a factor of two to higher and lower
values. These and other parameters of the magnetically
(``B'')-regulated star formation models are also listed in
Table~\ref{tab:SFmodels}. We note that although the true mass-to-flux
ratio depends on the geometry of the entire flux tube and cannot be
completely confined to a localized quantity, this only acts as a
first-order correction.

If a cell is magnetically subcritical (i.e., $\mu_{\rm cell}<1$), the
magnetic pressure is deemed strong enough to withstand gravitational
contraction, preventing any stars from forming within that cell. For
those cells that are magnetically supercritical (i.e., $\mu_{\rm
  cell}>1$), then star formation may be allowed to occur. However the
two other criteria introduced in \S\ref{S:denSF}, i.e., the cell is
resolved at the finest refinement level and $T<3000$~K, also must be
satisfied. In addition, we also limit the fraction of gas mass in a
cell that is turned into stars at a single timestep to be
$<0.5$. Given the minimum star particle mass, $m_{\rm \star,min}$,
this imposes an ``effective density threshold'' for the
magnetically-regulated star formation model. For this reason, we will
investigate magnetically-regulated models with $m_{\rm \star,min} =
10\:M_\odot$ (as in the density-regulated models), but also with
$m_{\rm \star,min} = 1\:M_\odot$. These choices correspond to
effective minimum threshold densities of $n_{\rm H,sf}= 3.55\times
10^{5}\:{\rm cm^{-3}}$ for $m_{\star, {\rm min}}=10\:M_{\odot}$ and
$n_{\rm H,sf}= 3.55\times 10^{4}\:{\rm cm^{-3}}$ for $m_{\star, {\rm
    min}}=1\:M_{\odot}$.

If all the above conditions are satisfied, then the star formation
process is allowed to occur at fixed efficiency per local free-fall
time, as described in \S\ref{S:denSF}. We will see that the
magnetically-regulated models with $m_{\star, {\rm min}}=1\:M_{\odot}$
can form significant numbers of star particles out of the stochastic
regime, but these masses should not be interpreted as being a
realistic assessment of the stellar initial mass function, since their
values depend on the size of the simulation timestep.

Overall there are four magnetically (``B'')-regulated runs (i.e.,
three choices of mass-to-flux threshold for $m_{\star, {\rm
    min}}=10\:M_{\odot}$ and one run with $m_{\star, {\rm
    min}}=1\:M_{\odot}$ at the fiducial mass-to-flux threshold) for
each of the noncolliding (``nocol'') and colliding (``col'')
simulation set-ups. The parameters of these star formation models and
simulations are also listed in Table~\ref{tab:SFmodels}.

\subsubsection{Star particle and star cluster dynamics}\label{S:stardyn}

Once the star formation criteria are met, mass is removed from the
cell and placed into a point-like star particle. These evolve as a
collisionless N-body system. However, these are not treated as
accreting sink particles, so they do not gain additional mass from the
gas, which we expect to be realistic due to the action of stellar
winds from the young stars.  The particles still interact with the gas
gravitationally via a cloud-in-cell (CIC) 
algorithm which maps the particle positions onto the grid. This limits
the closest distances between star particles to the grid resolution,
ultimately resulting in softer mutual gravitational interactions.  As
a result, small scale, i.e., internal, star cluster dynamics is not
expected to be well-modeled.  However, the early stages and larger
scales of the spatial and kinematic distribution of the stars, should
be more accurately followed.

Note also that our ability to follow the true internal dynamics of the
formed star clusters is limited by the fact that we do not fully allow
for the presence of a range of stellar masses, including both low-mass
and high-mass stars, or the presence of binary or higher order
multiple star systems. However, since our ability to accurately follow
the dynamical evolution of the star cluster is mostly limited by the
fact that gravitational forces are not well resolved below the grid
scale of the simulation, our focus is mostly on the global
distribution of stars in the simulations and the large scale spatial
and kinematic distributions of the stars in the clusters, e.g., low
order spatial mode asymmetries.

The current modeling also does not include feedback from the formed
star particles. A goal of a future paper is to include protostellar
outflow momentum feedback in these models, but at the moment the star
formation that results should be considered a baseline estimate in the
limit of zero feedback.

\section{Results}
\label{sec:results}

We perform analysis of each the simulations, comparing and contrasting
star formation models as well as non-colliding vs. colliding cases. In
particular, we discuss: morphology of the clouds and clusters
(\S\ref{sec:results-morph}); properties of star-forming gas
(\S\ref{sec:results-SFcells}); global star formation rates
(\S\ref{sec:results-SFR}); spatial clustering of stars
(\S\ref{sec:results-clust}); and star vs. gas kinematics
(\S\ref{sec:results-SFppv}).

\subsection{Cloud and Cluster Morphologies}
\label{sec:results-morph}

\begin{figure*}[htb]
\centering
\includegraphics[width=2\columnwidth]{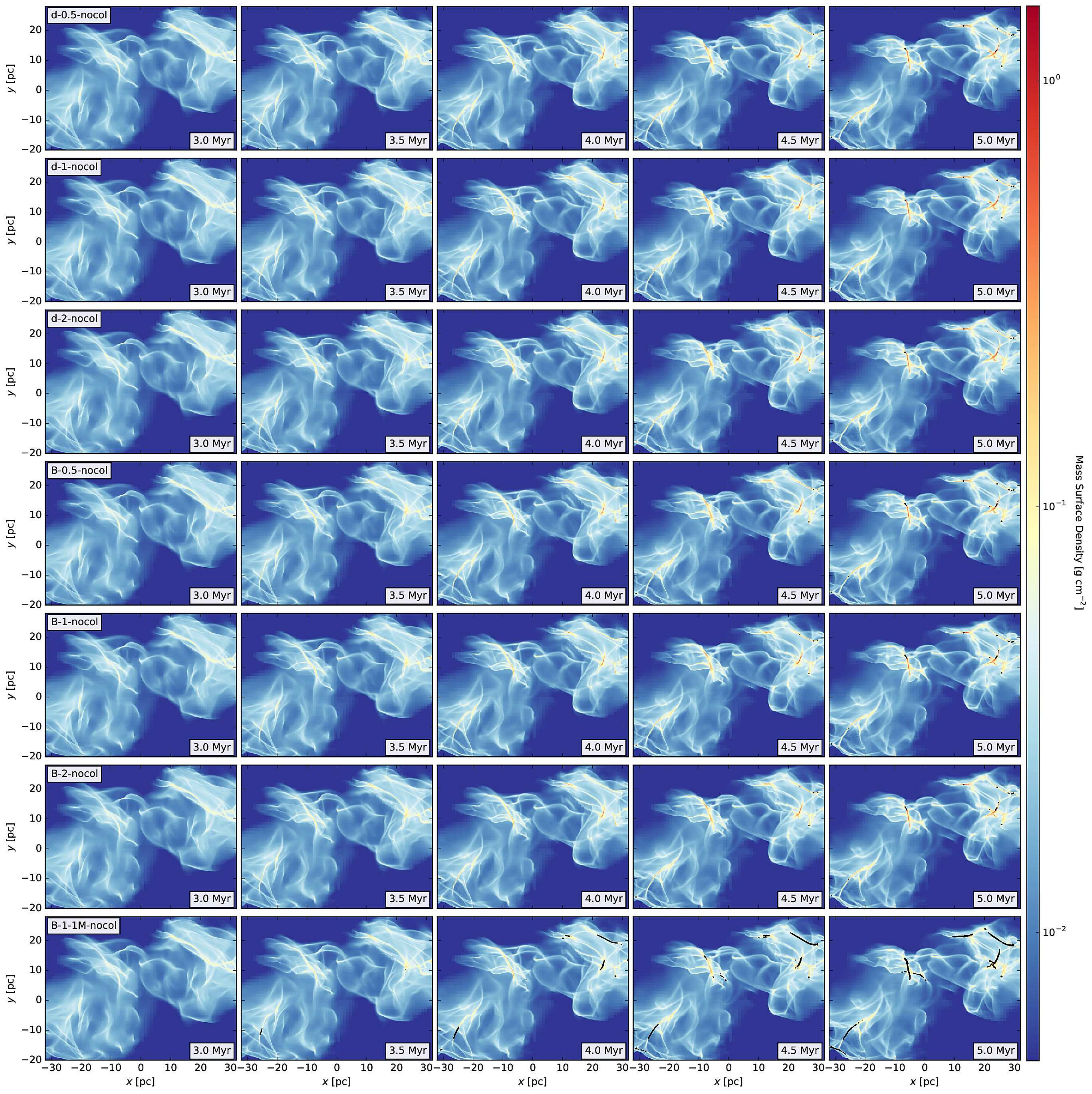}
\caption{
Time evolution of mass surface density, viewed along the $z$-axis, for
all non-colliding cases with star formation.  The top three rows
display the density-regulated SF runs, while the bottom four rows
display the magnetically-regulated SF runs, with labels given in the
left column.  Snapshots at 3.0, 3.5, 4.0, 4.5, and 5.0 Myr (left to
right) are shown.  Star particles are overplotted as black points.
\label{fig:sigma_SFnocol}}
\end{figure*}

\begin{figure*}[htb]
\centering
\includegraphics[width=2\columnwidth]{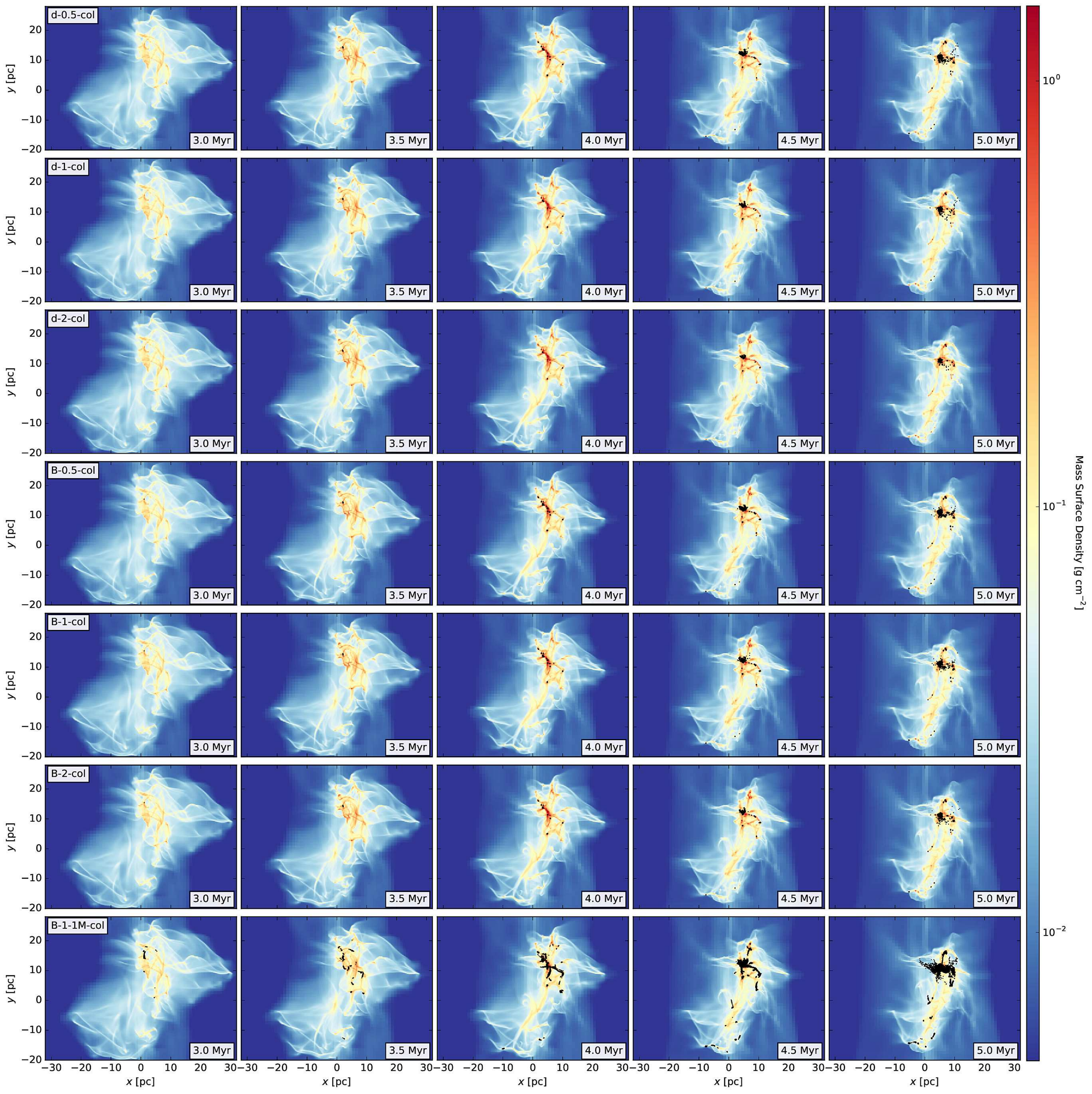}
\caption{
Time evolution of mass surface density, viewed along the $z$-axis,
i.e., perpendicular to the collision axis, for all colliding cases
with star formation. The top three rows display the density-regulated
SF runs, while the bottom four rows display the magnetically-regulated
SF runs, with labels given in the left column.  Snapshots at 3.0, 3.5,
4.0, 4.5, and 5.0 Myr are shown (left to right).  Star particles are
overplotted as black points.
\label{fig:sigma_SFcol}}
\end{figure*}

\begin{figure*}
\centering
\includegraphics[width=1\columnwidth]{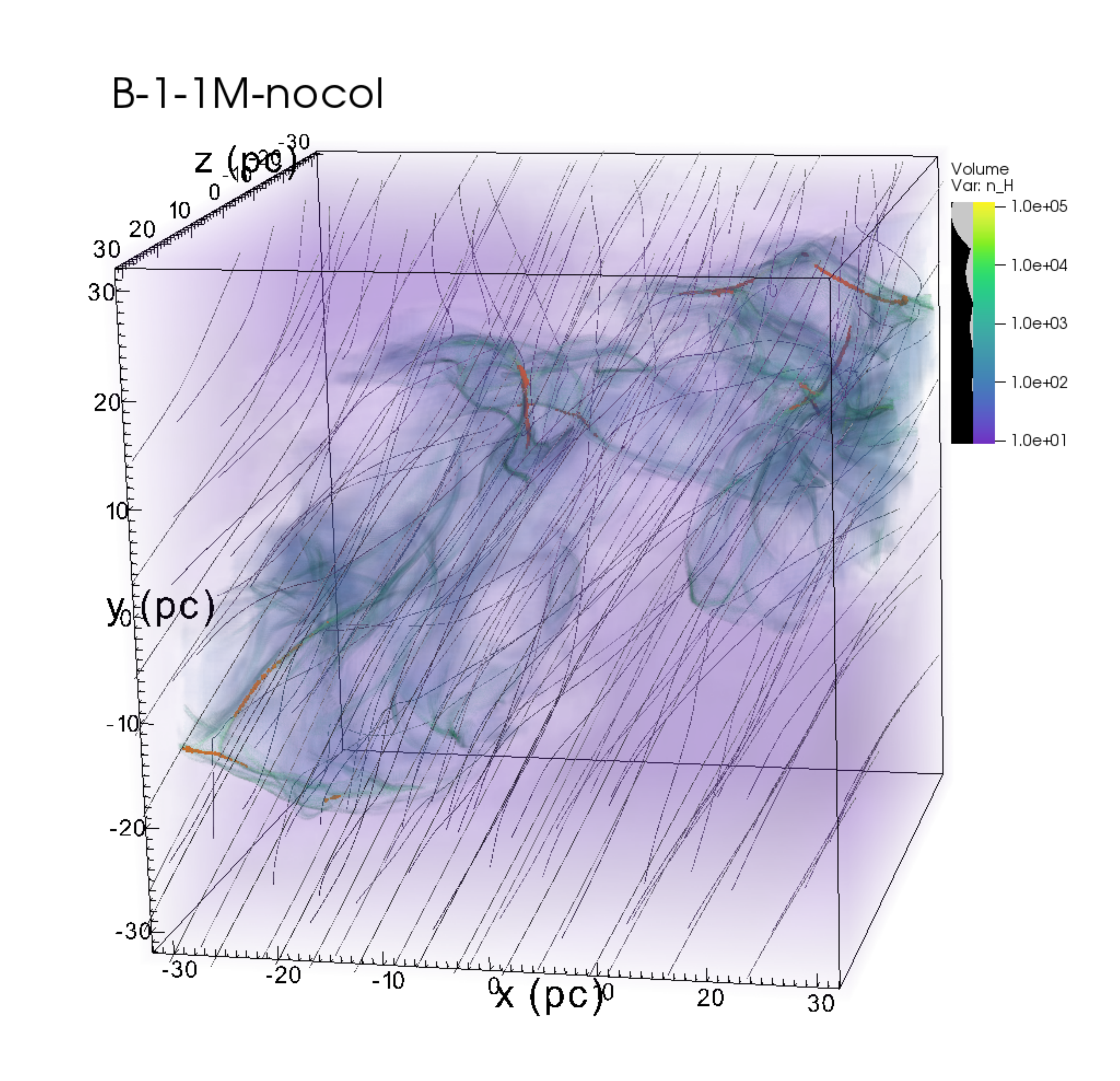}
\includegraphics[width=1\columnwidth]{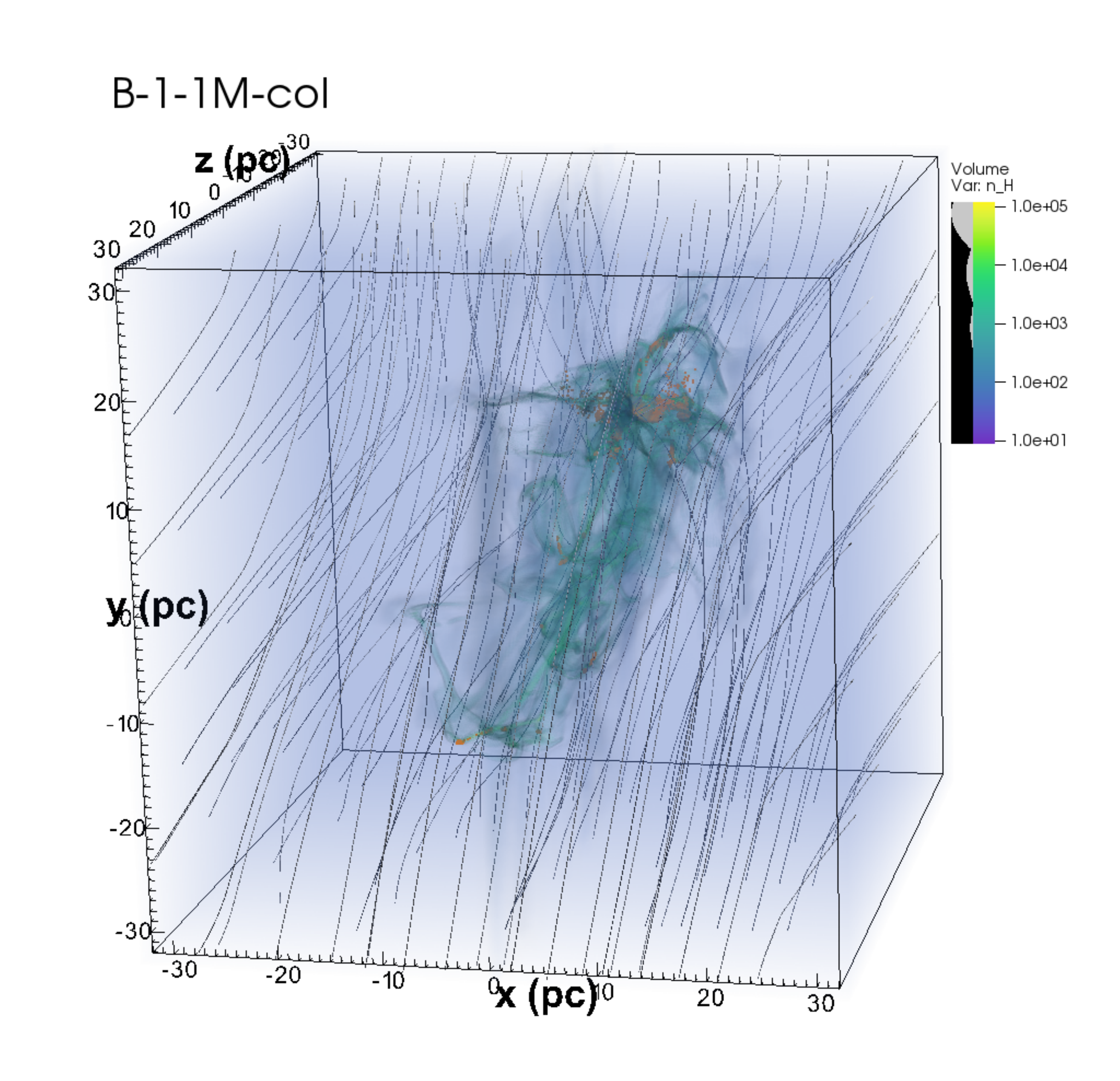}
\caption{
3D volume rendering of gas density shown together with magnetic field structure (streamlines) and star particles (orange points). Outputs at $t=5.0~{\rm Myr}$ are shown for the non-colliding (\textit{left}) and colliding (\textit{right}) models using the $m_{\rm \star,min}=1\:M_{\odot}$ magnetically regulated star formation routine. The transfer function is shown along with $n_{\rm H}$.
\label{fig:vol_rendering}}
\end{figure*}

The morphology of the gas and the stars are shown in
Figure~\ref{fig:sigma_SFnocol} for the non-colliding clouds and
Figure~\ref{fig:sigma_SFcol} for the colliding clouds. In the
non-colliding cases, the gas evolution is essentially identical, where
turbulent velocities and self-gravity create a network of relatively
slowly growing filaments with increasing differentiation in mass
surface density. Evolution is relatively passive and quiescent.  In
general, the onset of star formation takes place near $4.5\:{\rm Myr}$
for the $m_{\star, {\rm min}}=10~M_{\odot}$ cases and $3.5\:{\rm Myr}$
for the $m_{\star, {\rm min}}=1~M_{\odot}$ case.

Both density-regulated and magnetically-regulated models result in 
pockets of localized star formation concentrated at density peaks 
within filamentary structures. The slightly more populated network of 
filaments in the northeast region forms a higher number of 
(small) star clusters, but overall, star formation is scattered 
sparsely throughout both GMCs and remains relatively spatially 
isolated.  The magnetically-regulated models form clusters with a 
higher degree of elongation, i.e., following the axes of the natal 
filaments. There is also slightly more widespread star formation 
activity compared with the density regulated case. By $t=5.0~{\rm Myr}$, 
approximately 5-8 clusters have formed in the density-regulated cases, 
with the higher critical density models forming fewer clusters, whereas 
10-12 separate clusters have formed in the magnetically-regulated cases.

Differences are more pronounced in the $m_{\star, {\rm
    min}}=1~M_{\odot}$ case, as stars form in elongated clusters along
the filaments instead of the more localized spherical clusters as in
the density regulated case or the slightly eccentric clusters as in
the $m_{\star, {\rm min}}=10~M_{\odot}$ magnetically regulated
cases. 

However, one needs to bear in mind that the magnetically-regulated
models also involve an effective minimum density threshold as an
additional requirement for star formation. This threshold density
depends on the minimum star particle mass that is allowed in the model
via the requirement that no more than 50\% of the cell's gas mass can
be converted to a star particle (see \S\ref{S:magSF}). These effective
threshold densities are $n_{\rm H,\star min}= 3.55\times 10^{5}\:{\rm
  cm^{-3}}$ for $m_{\star, {\rm min}}=10~M_{\odot}$ and $n_{\rm
  H,\star min}= 3.55\times 10^{4}\:{\rm cm^{-3}}$ for $m_{\star, {\rm
    min}}=1~M_{\odot}$. Thus the variation in $m_{\star, {\rm min}}$
is a way of investigating how varying this effective density threshold
influences the resulting stellar population. Recall that for
star-forming gas, the star formation activity in lower density regions
is suppressed because the rate scales inversely with the local
free-fall time, i.e., SFR~$=\epsilon_{\rm ff} m_{\rm cell}/t_{\rm ff}
\propto n_{\rm H}^{3/2}$. Thus the overall SFR in these
magnetically-regulated models will depend on both the probability
distribution function (PDF) of densities of the gas above the
effective threshold density that achieves the magnetic criticality
condition. In the simple density-regulated models, the SFR will simply
depend on the PDF of densities above the threshold density.

Considering now the GMC collision cases (Figure~\ref{fig:sigma_SFcol}),
we see that they produce a much more active and dynamic environment
that leads to creation of much denser gas structures (see also Paper
II). At the interface of the colliding flows, a primary high-density
filamentary structure is formed.  This relatively compact, sheet-like
structure lies predominantly in the plane perpendicular to the
collision axis, with smaller filaments extending outward in various
directions. Mass surface densities of $\sim1~{\rm g\:cm^{-3}}$ are
reached much sooner compared to the non-colliding case.  This results
in earlier and more rapid star formation, generally beginning near
$3.0$ to $3.5\:{\rm Myr}$ for the $m_{\star, {\rm min}}=10~M_{\odot}$
cases and earlier than $3.0\:{\rm Myr}$ for the $m_{\star, {\rm
    min}}=1~M_{\odot}$ case of magnetically-regulated star formation.

In all such cases, the clusters form in the central colliding region,
at the peaks of filaments located in the primary filamentary
network. These sites often correspond with overdense clumps located at
filament junctions, potentially pointing toward star formation
triggered by filament-filament interactions on the smaller scale. By
$t=4.5~{\rm Myr}$, the individual star clusters have grown and merged
into one dominant star cluster located near
$(x,y)=(5{\rm~pc},10{\rm~pc})$, while stars continue to form from
dense clumps scattered throughout the post-shock colliding region.
This large cluster appears to contain multiple populations of smaller
star clusters that have merged together through a combination of
gravitational attraction and initial velocity inherited from the natal
gas of the collision.  The spatially separated clusters from earlier
times have grown in population and are moving toward the main cluster,
while a few smaller clusters are continuing to form along the
still-colliding dense filamentary gas. By $t=5.0$~Myr, the main
cluster (which has grown to a few thousand stars in the $m_{\star,
  {\rm min}}=10~M_{\odot}$ case and a factor of 10 higher in the
$1~M_{\odot}$ case) is co-located with the majority of the dense gas,
as more star clusters form in the vicinity. There exists a small
population of individual stars that
form in relative isolation and/or are dynamically ejected from the
denser regions.

The factor of two variations in $n_{\rm H,sf}$ do not greatly alter
the overall cluster morphology. However, there are small differences
in total cluster number as well as cluster size corresponding to the
density threshold, with increasing thresholds leading to reduced star
and cluster formation. 
The magnetically-regulated models exhibit slightly earlier star
formation, initializing just prior to $t=3.0$~Myr in each case, and a
higher number of clusters formed, which culminates in a larger central
cluster at late times compared with the density-regulated models.
Within these models, increasing values of $c_{1}$ result in reduced
star formation overall, though the locations where star formation is
centered do not change.

The \texttt{B-1-1M-col} model initiates star formation the earliest,
with a primary central cluster and 5-8 smaller clusters already formed
by $t=3.0~{\rm Myr}$.  Stars form in elongated structures directly
corresponding to the dense gas filaments similar, on small scales, to
that of the \texttt{B-1-1M-nocol} model.  By $t=4.0~{\rm Myr}$, stars
are present throughout the primary filament, still generally following
the filamentary structure of the gas, with smaller clusters forming
elsewhere throughout the colliding region. By $t=5.0~{\rm Myr}$, the
primary central cluster has grown directly as well as from
gravitational interactions with the nearby clusters. Outlying clusters
have continued to increase in size and number.

Within dense filaments, the B-field is generally aligned perpendicular
to the filament axis (see Paper II). Qualitatively, the mass-to-flux
is expected to be highest at the density peaks, locally, but is
expected to decrease when the entire flux tube is taken into account
due to the lower-density environment surrounding the filaments. 
We note also that although these are ideal MHD simulations, some
numerical diffusion of flux is expected to occur that may influence
the star formation activity. The effects of modeling non-ideal MHD
processes will be explored in a future paper in this series.

Figure~\ref{fig:vol_rendering} shows the combined 3D structure of the gas 
density, magnetic field geometry, and star particles. The \texttt{B-1-1M-nocol}
and \texttt{B-1-1M-col} models are compared at the same time, $t=5.0~{\rm Myr}$, 
revealing the denser and more compact structure created in the GMC collision. 
These figures show the contrasting global morphology of the gas and stellar structure. Detailed analysis of various aspects of these properties is performed
in subsequent sections.

\begin{figure*}
\centering
\includegraphics[width=1\columnwidth]{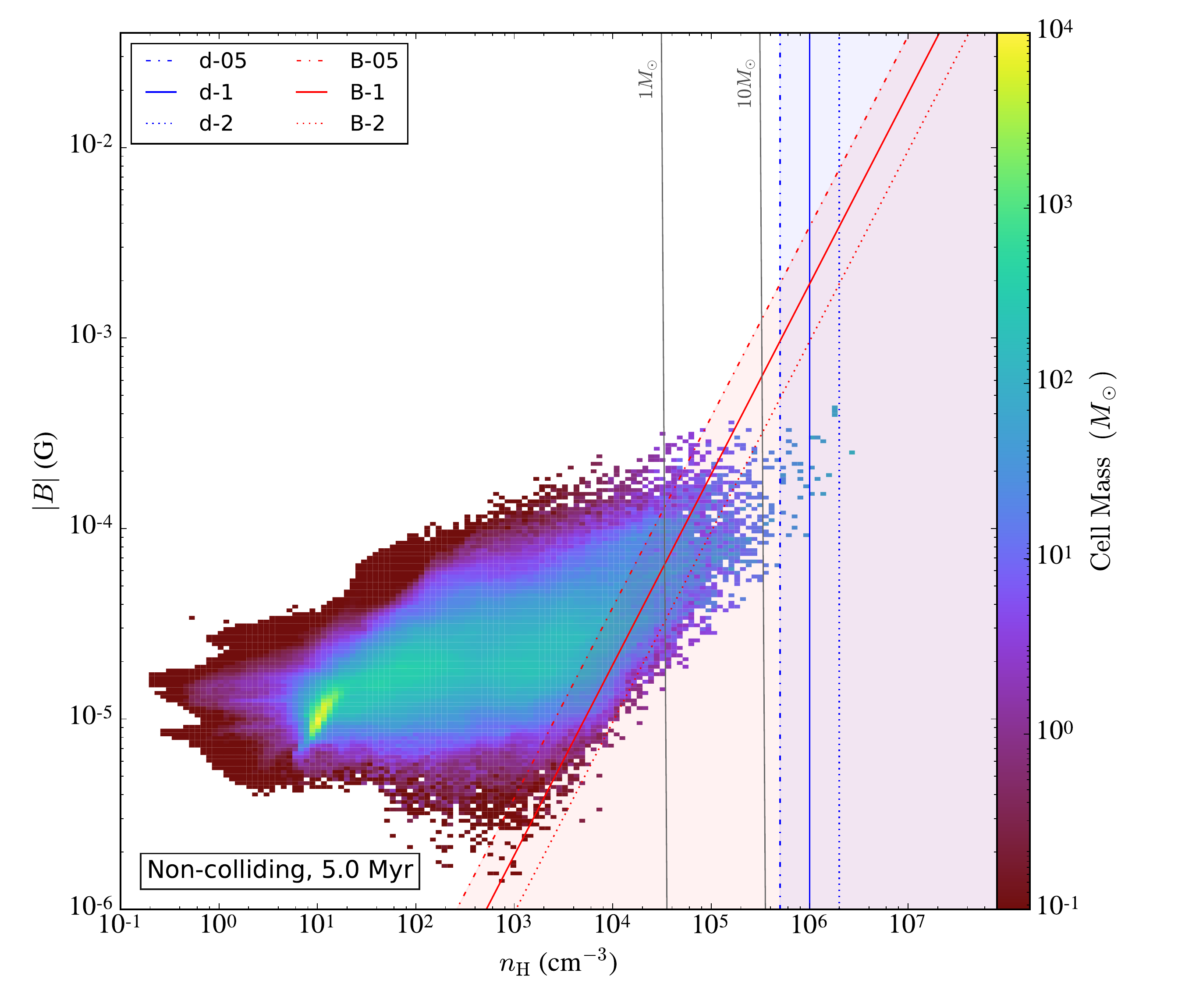}
\includegraphics[width=1\columnwidth]{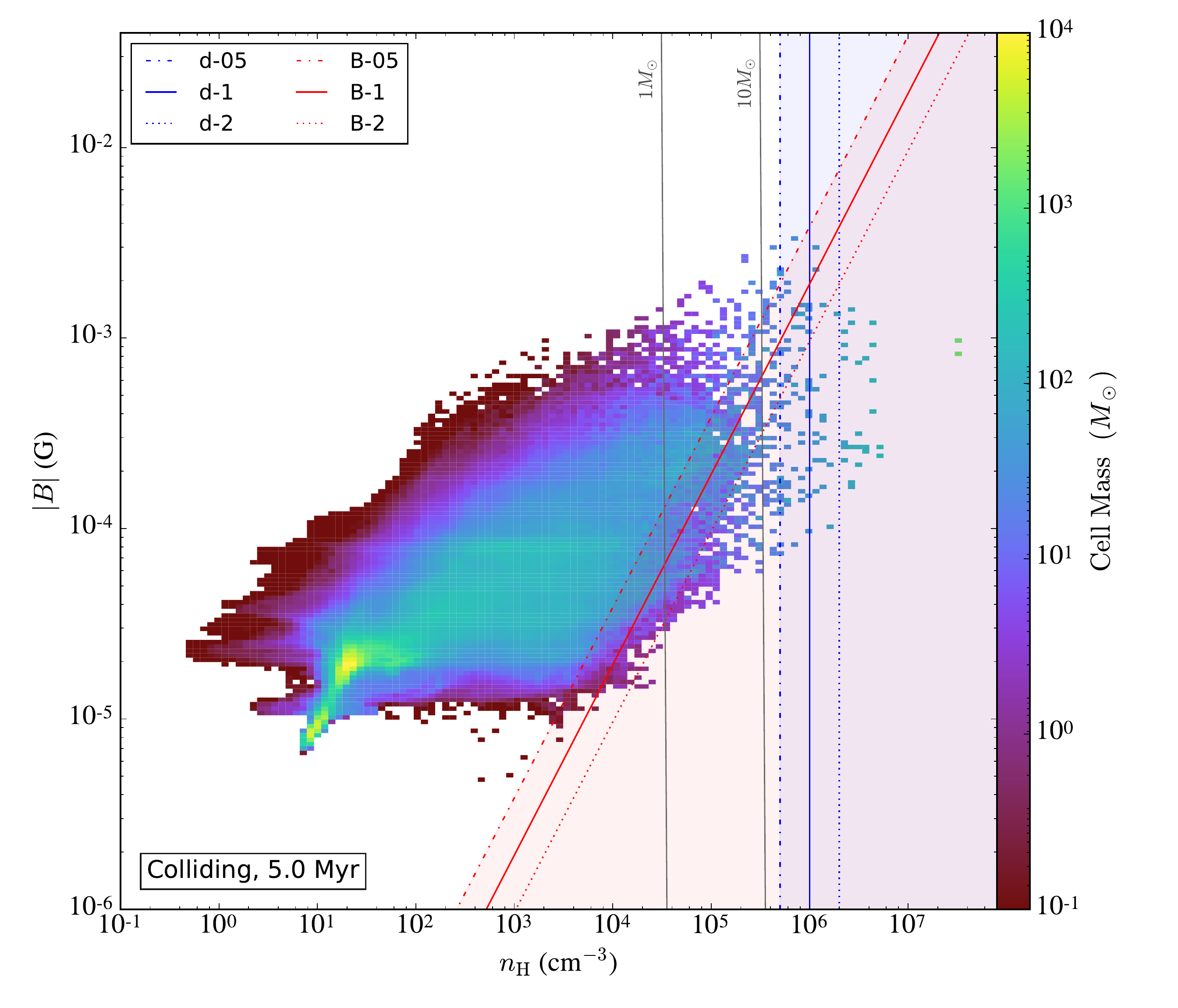}
\caption{
Phaseplots in $n_{\rm H}$ vs. $B$ space of the non-colliding (left
panel) and colliding (right panel) cases at 5~Myr for non-star-forming
models. The blue lines represent the star formation density threshold
$n_{\rm H,sf}$ for the various critical densities in the
density-regulated star formation routine. The blue-shaded region
represents the regime in which star particles would form from these
methods.  The red lines represent the various mass-to-flux thresholds
for the magnetically-regulated star formation routine, with the
red-shaded region showing conditions needed for star particle
formation.  The black lines show the effective minimum densities due
to the 50\% mass limitation for two values of $m_{\rm \star, min}$.
\label{fig:phase_noSF}}
\end{figure*}

The different star formation models 
can be directly compared via $|B|$ vs. ${n_{\rm H}}$ phaseplots
(Figure~\ref{fig:phase_noSF}). 
The critical thresholds for each model are plotted over phaseplots in 
respective non-colliding and colliding runs without star formation
(i.e., fiducial runs from Paper II extended to $t=5.0~{\rm Myr}$). 
In this manner, the total mass along with various properties of the 
gas affected by each star formation model can be estimated.
The colliding case forms regions of overall higher density and
magnetization, both enhanced by approximately an order of magnitude.
At a given density, gas in the colliding case generally contains
stronger field strengths due to the nature of the compressive
flows. The star formation thresholds for both the density-regulated
and magnetically-regulated star formation routines are overplotted as
blue and red lines, respectively.

The density-regulated star formation regime affects a greater total
gas mass in the colliding case. As the critical density threshold
decreases, the number of affected cells in both scenarios increases,
leading to increased star formation regardless of magnetic field
strength.

As the threshold for mass-to-flux ratio is lowered, a similar pattern
of increasing star formation occurs. 
Key differences from the density-regulated models become
apparent as star formation is now allowed to occur in regimes of low 
density, super-critical gas and is inhibited in high-density, 
sub-critical gas. Overall, the various models primarily create stars 
from the same gas, though narrow regimes exist in which stars form 
exclusively within certain routines.
As discussed above, in these
magnetically-regulated models, the effective minimum density
thresholds for star formation provide an additional bound.  In the
$m_{\star, {\rm min}}=10~M_{\odot}$ models, much of the gas in the
colliding cases--and even more so in the non-colliding cases--is
limited by this effective density threshold. 
The $1~M_{\odot}$ cases allow star formation from a larger amount of
locally super-critical gas. However, we will see below that the
overall mass of stars formed by 5~Myr in the colliding case depends
only weakly on this choice.

\subsection{Properties of Star-forming Gas}
\label{sec:results-SFcells}

\begin{figure*}[htb]
\centering
\includegraphics[width=2\columnwidth]{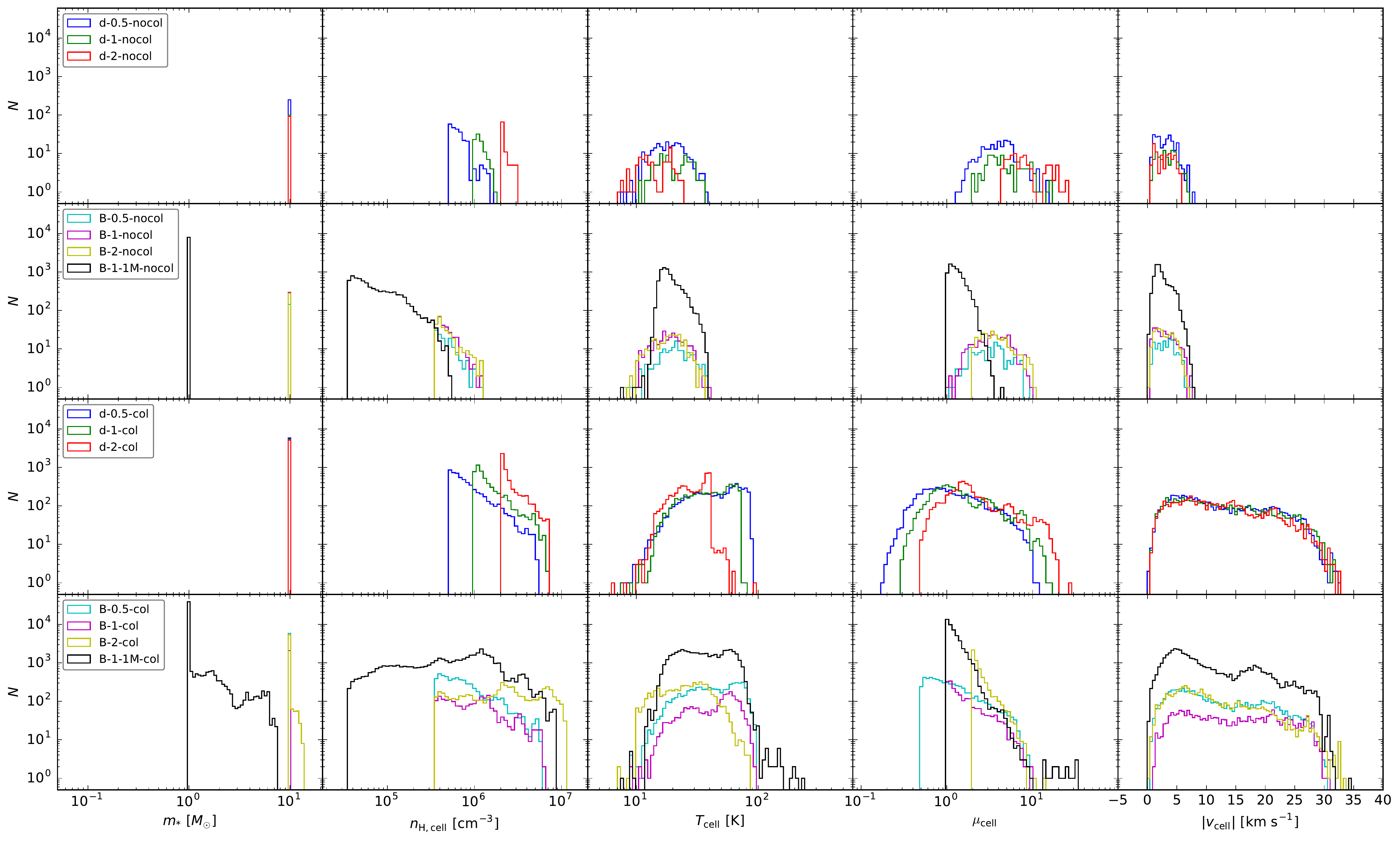}
\caption{
Masses of young stars and properties of star-forming gas
cells. Histograms of (left to right) stellar mass, number density of
the cells that form stars, gas temperature of such cells, normalized
mass-to-flux ratio of such cells and velocity magnitude of such cells
are shown for the non-colliding (top two rows) and colliding (bottom
two rows) cases. The thresholds for critical densities and effective
minimum densities can be seen as cutoffs for the density-regulated and
magnetically-regulated star formation models, respectively. The
critical mass-to-flux values can also be seen for the
magnetically-regulated models.
\label{fig:SFcells_hist}}
\end{figure*}

We examine the masses of young stars and properties of their
progenitor gas cells, just before a star particle is
created. Figure~\ref{fig:SFcells_hist} displays the cumulative histograms
over 5~Myr of the stellar masses and key properties of the
star-forming gas.

For the non-colliding cases, the stars form strictly at their
threshold masses, indicating purely stochastic star formation is
occurring. The density-regulated models form approximately 100-250
stars each, with higher critical density thresholds resulting in fewer
stars.  The distributions of cell densities peak at the thresholds of
0.5, 1, and $2 \times 10^{6}\:{\rm cm^{-3}}$ for the respective models
and extend above the cutoffs by factors of a few. Gas temperatures
range from 6 to 40~K, averaging approximately 10 to 20~K with higher
density thresholds resulting in slightly lower temperatures. The local
normalized mass-to-flux ratio of the star-forming cells in these
models is supercritical by factors of few for \texttt{d-0.5-nocol} to
few tens for \texttt{d-2-nocol}. The velocities of these cells are
generally a few ${\rm km~s^{-1}}$, consistent with values expected
from decaying turbulence in the self-gravitating GMCs.

The magnetically-regulated non-colliding models exhibit slightly
higher numbers of stars (few hundred for the $m_{\star, {\rm
    min}}=10~M_{\odot}$ models).  Across these three models,
distributions for density have peaks at the cutoff of $\sim
3\times10^{5}\:{\rm cm^{-3}}$, temperatures primarily near 20~K
(slightly above equilibrium), $\mu_{\rm cell}$ near 3-5 (slightly
supercritical), and $|v|$ near 2~${\rm km\:s^{-1}}$ (supersonic but
consistent with decay of the initial turbulence). There exist slight
trends of increasing density and decreasing temperature as the
thresholds for $\mu_{\rm cell}$ increase between models. For the
$m_{\star, {\rm min}}=1~M_{\odot}$ model, approximately 20 times more
star particles are created. 
The distribution of cell densities also exhibit a cutoff at the
effective minimum density of $\sim 3\times10^{4}\:{\rm cm^{-3}}$ in
this model, with the spread of densities reaching up to a factor of 20
higher. The temperature and velocity distributions exhibit similar
peaks and spreads as their higher minimum stellar mass
counterparts. However, the cells generally have a lower $\mu_{\rm
  cell}$ near 1 (magnetically critical) when stars are formed. This
suggests that the condition for criticality is reached before the cell
density grows to a point at which it can produce more massive stars in
a given timestep and so stays in the stochastic limit.

The colliding models with density-regulated star formation produce
$\sim 6\times10^{3}$ stars, over 20 times the number formed from the
non-colliding models over the same time period. The density
distributions are similar to the non-colliding cases, peaking at the
cutoffs, but exhibit an increased spread with cells reaching densities
higher by factors of a few. The gas temperatures are also higher,
averaging near 30-40~K but with a few cells reaching $\sim 90~{\rm
  K}$. Temperatures are generally lower for the higher-density cutoff
models, but all peak at temperatures higher than equilibrium, likely
due to shocks produced throughout the primary colliding region. The
collision also produces high-density gas at a wide range of $\mu_{\rm
  cell}$, ranging from a few times subcritical up to $\sim 20$ times
supercritical. The distributions peak near the value for magnetic
criticality, with lower higher-density cutoff models corresponding
with higher values of $\mu_{\rm cell}$. 
Cell velocity distributions
are nearly identical, peaking near 10-20~${\rm km\:s^{-1}}$.

For the magnetically-regulated colliding models with
$m_{\star,{\rm min}}=10~M_{\odot}$, a similar star particle count
is seen, exceeding their respective non-colliding models by factors of
10 to 20. The \texttt{B-2-col} model also forms some stars outside of
the stochastic regime, as masses of $\sim {\rm 12-13}M_{\odot}$ are
created. It is important to recall that the expected mass of the star
particle to be created depends on the local SFR in the cell, i.e., on
$\epsilon_{\rm ff}$ and the cell density, but also on the timestep of
the simulation. Thus the presence and mass distribution of these
higher mass star particles should not be over-interpreted. The
presence of stars outside the stochastic regime simply indicates that
some very high density, high SFR cells are present, and this is
confirmed in the plots showing the density distributions, with some
densities up to $n_{\rm H}=10^7~{\rm cm^{-3}}$. We note that
star-forming cells can also have higher temperatures near 30 to 40~K,
perhaps indicating creation of the dense gas in shocks, but recall
that the star formation sub-grid model does not assess the degree of
gravitational instability in the gas. The star-forming cells show a
concentration of gas at the minimum magnetic criticality cutoff.
They also exhibit generally higher velocities indicating strong 
turbulence and/or bulk motion associated with the GMCs.

The \texttt{B-1-1M-col} model has the greatest total number of stars
formed, i.e., $\sim 5\times10^{4}$ and forms a range of stellar masses
up to $\sim 7~M_{\odot}$ (but again this mass function should not be
expected to be compared to a real IMF, rather being simply the way the
model ensures the total mass of stars formed is correct given the
model parameters). 
Cell number densities range from $\sim 3\times10^{4}$ to $\sim
9\times10^{6}\:{\rm cm^{-3}}$. Temperatures are near 40~K and $\mu_{\rm
  cell}$ reaches a few tens but increases in cell number towards the
critical value cutoff of 1. Velocities exhibit a similar trend as the
$10~M_{\odot}$ models, showing high levels of turbulence, bulk
motion,  and/or infall to the primary cluster.

\subsection{Star Formation Rates and Efficiencies}
\label{sec:results-SFR}

\begin{figure*}[htb]
\centering
\includegraphics[width=1.8\columnwidth]{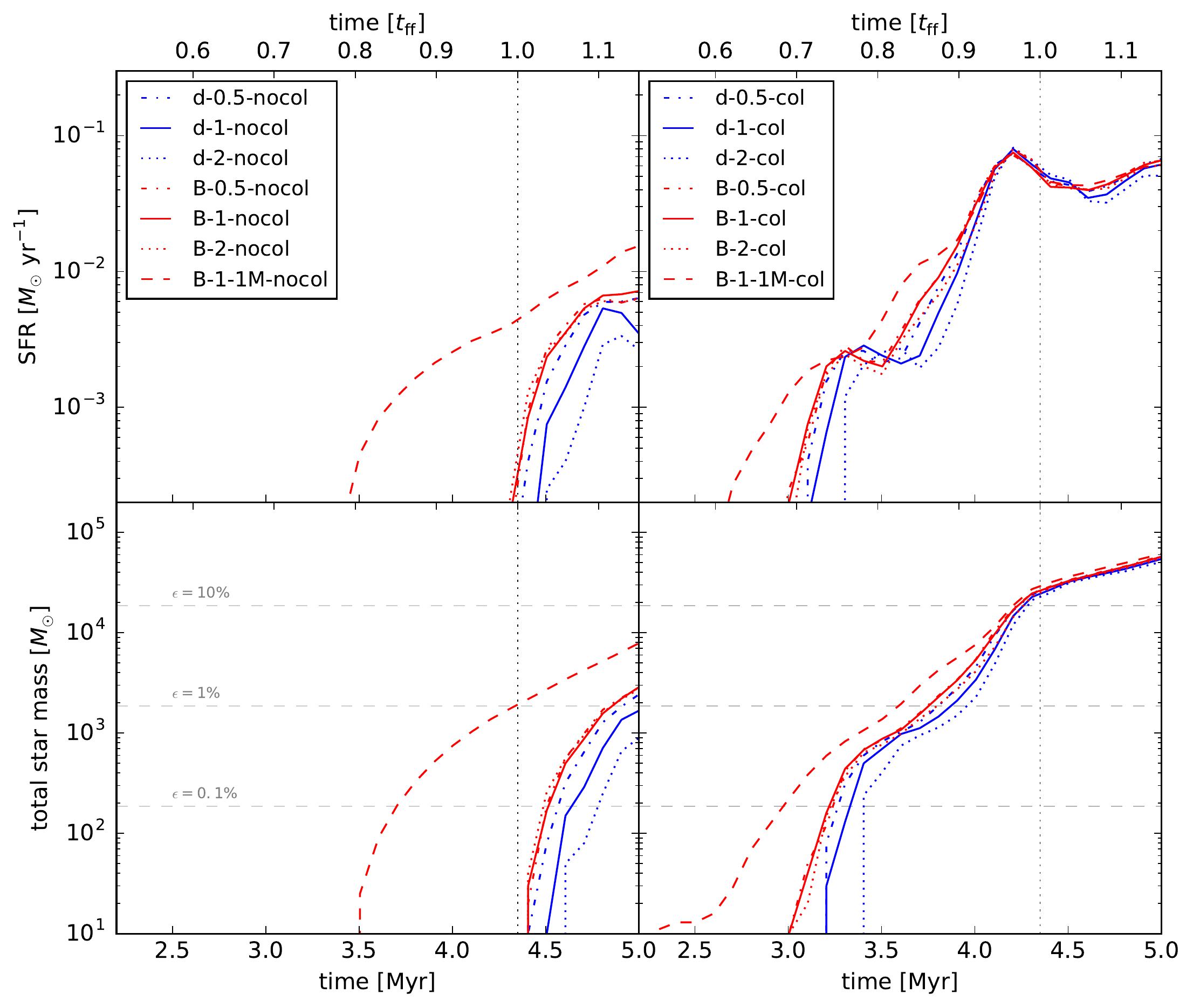}
\caption{
{\it Top row:} 
The star formation rates over time. The non-colliding cases are shown
in the left column, while the colliding models are in the right
column. Desity-regulated star formation models are traced with blue
lines; magnetically-regulated models with red lines. The vertical
dotted line shows the time after 1 free-fall time of the initial GMCs
(see top axis).
{\it Bottom row:} The total mass of stars formed versus time. The total star
formation efficiency, shown for $\epsilon$=0.1\%, 1\%, and 10\%, is
normalized relative to the total initial mass within the two GMCs
($1.86\times 10^{5}\:M_{\odot}$). 
\label{fig:SFR}}
\end{figure*}

The star formation rate (SFR) and overall star formation efficiency
(SFE) of molecular clouds are important quantities that help determine
the global galactic star formation process.  The time evolution of
these quantities for both non-colliding and colliding cases, for each
star formation model, is shown in Figure~\ref{fig:SFR}.

The SFR is calculated as the time derivative of the total mass of the
star particles. The efficiencies are determined by normalizing the
total stellar mass by the combined gas mass of the two original
GMCs. The evolution of these quantities is measured in simulation time,
as well as relative to the freefall time of the initial
GMC density ($t_{\rm ff}=4.35$~Myr).

In the non-colliding density-regulated cases, star formation initiates
shortly after $1 t_{\rm ff}$, i.e., at approximately $t=4.4$~Myr. The
higher-cutoff density models form stars at slightly later times
corresponding to when the critical density is achieved. At a given
time, SFR and SFE vary by factors of a few between models. Over the 
course of the next 0.5~Myr (until simulation completion), the SFRs 
increase to $\sim (3-6)\times10^{-3}\:M_{\odot}\:{\rm yr^{-1}}$ and 
then generally level off. Total stellar masses of $\sim
9\times10^{2}-3\times10^{3}\:M_{\odot}$ are created, corresponding to
$\epsilon \approx 1\%$ by $1.15~t_{\rm ff}$.

For the magnetically-regulated cases with $m_{\rm
  \star,min}=10\:M_\odot$, star formation also starts after $1 t_{\rm
  ff}$, and evolves in a similar manner as the density-regulated
models except with slightly higher SFRs and efficiencies. The
differences between these models is also much smaller, as the three
$10~M_{\odot}$ magnetically-regulated models reach about
$6\times10^{-3}\:M_{\odot}\:{\rm yr^{-1}}$ by $1.1~t_{\rm ff}$ and
then level off. The SFEs also reach and slightly exceed 1\% by the
simulation end time. The \texttt{B-1-1M-nocol} case exhibits the most
dissimilar behavior of the non-colliding models, initiating star
formation approximately 1~Myr earlier and reaching
$4\times10^{-3}\:M_{\odot}\:{\rm yr^{-1}}$ and $\epsilon=1\%$ by $1
t_{\rm ff}$. By 5~Myr, the SFR and SFR are approximately 2-3 times
greater than the other magnetically-regulated models.

The above trends are mostly likely caused by the fact that all these
star formation models have effective density thresholds that need to
be met to allow star formation to proceed, even the
magnetically-regulated models (see \S\ref{S:magSF}). These thresholds
decrease monotonically as we consider the density-regulated models,
then the B-(0.5,1,2)-nocol models, and then the B-1-1M-nocol
model. The simulations are in a regime in which the total SFR and
eventual total SFE are set mostly by the fraction of gas in the GMCs
that can meet these density threshold criteria. For the particular
$B$-field strengths in these simulations (i.e., 10$\:{\rm \mu G}$),
the choice of the magnetic threshold parameter does not play a
significant role in setting the SFR.

The colliding cases produce much higher SFRs and SFEs during their
evolution. The density-regulated models begin forming stars 
at a rapid pace shortly after $t=3.2$~Myr, with higher density 
thresholds slightly delaying the onset of star formation. There is some
oscillation in the growth of the SFRs, but overall it increases from
onset until $t_{\rm ff}$ near $0.08~M_{\odot}\:{\rm yr^{-1}}$ and 
then levels off through the culmination of the simulations.
Star formation efficiencies reach 1\% by 3.7 to 4~Myr and more than 20\% by
$t_{\rm ff}$. While the early behavior differs slightly in time between
the density-regulated models, they appear to converge at later times.

The $m_{\rm \star,min}=10\:M_{\odot}$ magnetically-regulated colliding
cases show very similar results to each other throughout the whole
evolution, which indicates that the SFR is not limited by the
mass-to-flux thresholds in this simulation set-up. Indeed, these
models also converge with the density-regulated models by about
4~Myr. 
\texttt{B-1-1M-col} starts forming stars at the earliest
times, but it also shows convergence in SFR by about 4~Myr. 
These results indicate that the SFR is in fact not limited by the density
threshold criteria either. In the GMC-GMC collision the SFRs appear to
be set by the creation of structures that can place gas at densities
greater than any of the threshold densities, after which, even with
$\epsilon_{\rm ff}=0.02$, it is turned quite efficiently into stars.

It is important to note that stellar feedback has not yet been
included in our star formation models. Our current treatment may be a
good approximation for initial 
SFRs, but mechanisms such as protostellar outflows that become 
important during formation, and ionization, winds and radiation pressure 
from massive stars soon after, will likely result in reduced SFRs.

\subsection{Spatial Clustering}
\label{sec:results-clust}

We investigate various quantitative metrics for spatial structure
of the star clusters formed in our simulations. Global star and gas 
properties of the primary clusters are measured and the
\textit{angular dispersion parameter} (ADP) and \textit{minimum
spanning tree} (MST) methods are used to analyze cluster substructure.
The ADP is sensitive to angular substructure at chosen radii, while 
the MST determines the degree of overall centrally concentrated 
clustering.

In order to define the primary cluster within a given model, we use
the \textit{Density-based spatial clustering of applications with
noise} (DBSCAN) algorithm \citep{Ester_ea_1996}. This density-based
clustering algorithm is applied to our projected star particle data
and the median particle position of the highest population cluster is
used as our initial cluster center. A circular aperture with initial
radius of 0.4~pc is centered at this point, and a new center is 
determined by finding the center of mass using stars included only 
within this aperture. This process is repeated with aperture radii 
iteratively decreasing by factors of two, down to length scales of
0.1~pc. The ADP is found for the primary cluster using each of these
defined centers, while the MST is found for the entire domain.

\vspace{10mm}

\subsubsection{Global structure of primary cluster}
\label{S:clust-props}

\begin{figure*}[htb]
\centering
\includegraphics[width=2\columnwidth]{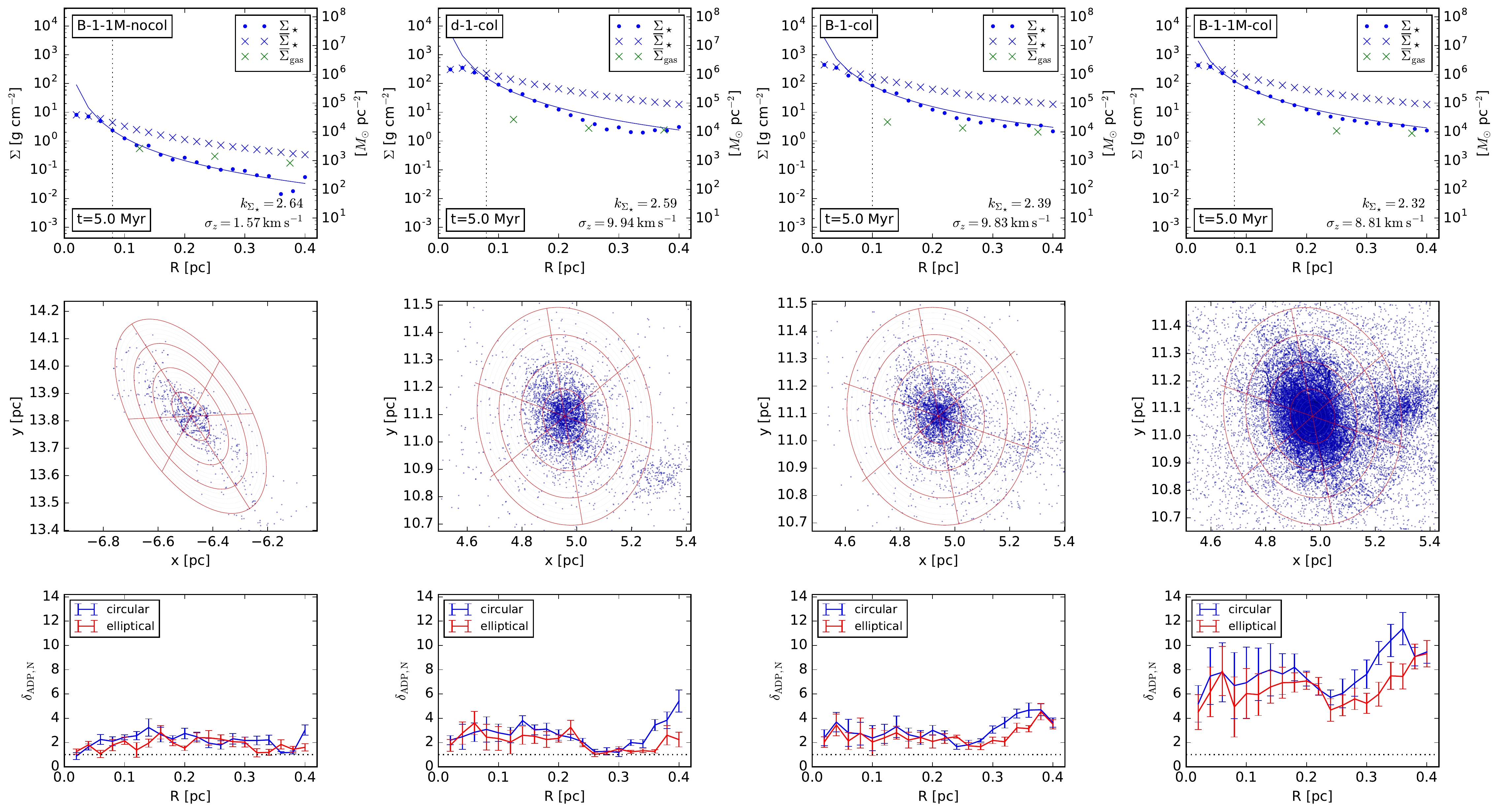}
\caption{
{\it Top row:} Mass surface densities as a function of radius for
stars ($\Sigma_{\star}$) and gas ($\Sigma_{\rm gas}$) composing the
primary clusters in various models at $t=5.0$~Myr. From left to right
the models shown are \texttt{B-1-1M-nocol}, \texttt{d-1-col},
\texttt{B-1-col} and \texttt{B-1-1M-col}. Blue circles indicate
$\Sigma_{\star}$ calculated locally in each circular annulus, while
blue crosses show the enclosed average quantity, $\overline{\Sigma}_*$.  A
power law fit to $\Sigma_\star$ is shown as the blue line and the
resulting exponent $k_{\Sigma_\star}$ is displayed. The black dotted
line denotes the position of the stellar half-mass radius,
$R_{1/2}$. The green crosses indicate the average enclosed gas mass,
$\overline{\Sigma}_{\rm gas}$.  The 1-D velocity dispersion of the stars,
$\sigma_{z,1/2}$, is calculated for the cluster defined by the
half-mass radius.
{\it Middle row:} 
Visualization of the primary clusters. The star particles are shown as
blue points, while the elliptical annuli and sectors used to calculate
the angular dispersion parameter, $\delta_{\rm ADP,N}$, are overlaid
in gray. Note that, for clarity, circular annuli that are used in
calculating both the cluster profiles (top row) and for one version of
the $\delta_{\rm ADP,N}$, are not displayed. Each sector is outlined
in red, while the annuli are shown as gray ellipses with every fifth
annulus highlighted in red.
{\it Bottom row:} $\delta_{\rm ADP,6}$ vs. radius from the chosen
cluster center. The values are calculated using both circular and
elliptical annuli and averaged over twenty $3^{\circ}$ rotations of
orientation of the 6-sector pattern. The error bars depict the
standard error of the mean.  A purely random azimuthal distribution of
particles is indicated by the dotted line at $\delta_{\rm ADP,6}=1$.
\label{fig:ADP}}
\end{figure*}

We measure global structural properties of the primary star clusters
created in the simulations. Figure~\ref{fig:ADP} shows results for
these clusters that have formed by $t=5~{\rm Myr}$ in four models:
\texttt{B-1-1M-nocol}, \texttt{d-1-col}, \texttt{B-1-col}, and
\texttt{B-1-1M-col}.  Due to relatively sparse particle density, the
clusters in the non-colliding $m_{\rm \star,min}=10\:M_{\odot}$ models
were not included in this analysis. Also note that while elliptical
annuli are displayed (see ADP discussion in Section~\ref{subsec:ADP}),
global properties for cluster structure are calculated using circular
annuli.

The top row of Figure~\ref{fig:ADP} shows the mass surface density of
stars locally within each annulus, $\Sigma_{\star}$, the enclosed
average mass surface density of stars, $\overline{\Sigma}_{\star}$,
and the enclosed average mass surface density of the gas,
$\overline{\Sigma}_{\rm gas}$.  For the non-colliding case,
$\overline{\Sigma}_{\star}$ ranges from about $10^{3}$ to $5\times
10^{4}~{\rm M_{\odot}\:pc^{-2}}$, whereas the colliding cases form
clusters with approximately $10^{5}$ to $10^{6}~{\rm
  M_{\odot}\:pc^{-2}}$. It can be seen that $\Sigma_{\star}$ falls off
quickly, reaching $\sim10^{2}$ and $\sim10^{4}~{\rm
  M_{\odot}\:pc^{-2}}$ by $R=0.4~{\rm pc}$ for non-colliding and
colliding cases, respectively. We find best fit power law profiles of
the form:
\begin{equation}
\Sigma_{*}(R) = A \left( \frac{R}{0.2~{\rm pc}}\right)^{-k_{\Sigma_*}},
\end{equation}
where $A$ is a normalization factor, $R$ is the distance from cluster
center, and $k_{\Sigma_\star}$ is the power law exponent.
$k_{\Sigma_\star}$ is found to be $\sim2.3$ to 2.6 for these primary
clusters.

We denote a half-mass radius, $R_{1/2}$, as the radius within which
half of the total mass of the cluster out to $R=0.4~{\rm pc}$ is
contained. Within our chosen clusters, $R_{1/2}\simeq0.08$-0.1~pc.
Within $R_{1/2}$, the total enclosed stellar masses are $4.0\times
10^{2}\:M_{\odot}$, $2.3\times 10^{4}\:M_{\odot}$, $2.3\times
10^{4}\:M_{\odot}$, and $2.2\times 10^{4}\:M_{\odot}$ for models
\texttt{B-1-1M-v00}, \texttt{d-1-v10}, \texttt{B-1-v10}, and
\texttt{B-1-1M-v10}, respectively. This shows that the properties of
the primary cluster in the colliding simulations are not much affected
by the choice of star formation subgrid model. Within $R_{1/2}$, the
averaged stellar mass surface density is
$\overline{\Sigma}_{*,1/2}=2.1\times 10^{4}\:{\rm M_{\odot}~pc^{-2}}$
($4.4\:{\rm g~cm^{-2}}$) for the non-colliding case and $1.1\times
10^{6}\:{\rm M_{\odot}~pc^{-2}}$ ($2.28\times10^{2}\:{\rm
  g~cm^{-2}}$), $8.0\times 10^{5}\:{\rm M_{\odot}~pc^{-2}}$
($1.7\times10^{2}\:{\rm g~cm^{-2}}$), and $1.06\times 10^{6}~{\rm
  M_{\odot}\:pc^{-2}}$ ($2.22\times10^{2}\:{\rm g~cm^{-2}}$) for the
respective colliding GMC models.  The respective gas masses are
$53\:M_{\odot}$, $5.7\times10^2\:M_{\odot}$,
$7.7\times10^2\:M_{\odot}$, and $4.5\times10^2\:M_{\odot}$.  Note that
the $\sim0.1~{\rm pc}$ scales are barely resolved in our simulation,
so only the average enclosed gas masses are measured.  At this stage,
the cluster in the non-colliding case has $\Sigma_{\rm \star} <
\overline{\Sigma}_{\rm gas}$. The colliding cases all have
$\Sigma_{\rm \star} > \overline{\Sigma}_{\rm gas}$.

We compare these cluster properties with those of observed young
clusters \citep[see, e.g., Figure 1 of][]{Tan_ea_2014}.  The cluster
formed in the non-colliding simulation \texttt{B-1-1M-v00} that has
$\overline{\Sigma}_{\star} \simeq 2\times10^{4}~M_{\odot}\:{\rm
  pc^{-2}}$ ($4.4\:{\rm g~cm^{-2}}$) is much denser than any known
young cluster of comparable mass (i.e., with
$M_{*,1/2}\lesssim1000\:M_{\odot}$).  The colliding simulations
produce more massive clusters and these are also seen to have much
higher mass surface densities (by more than a factor of ten) at their
half-mass scale than the densest known Galactic clusters, such as the
Arches or Westerlund 1. We note that stellar feedback is not currently
included within our simulations and expect the implementation of
protostellar outflow feedback, planned in a future paper, to result in
a reduction of $\overline{\Sigma}_{\star}$.

The 1-D velocity dispersion, $\sigma_{z,1/2}$, is calculated for the
stars seen to be within projected radii of $R_{1/2}$.  The cluster
formed in the non-colliding case has $\sigma_{z,1/2}=1.57\:{\rm
  km\:s^{-1}}$, while those formed from GMC collisions have much
higher values of 9.94, 9.83, and $8.81\:{\rm km\:s^{-1}}$,
respectively for these clusters shown in Figure~\ref{fig:ADP} (left to
right).  More detailed kinematic analysis is performed in
Section~\ref{sec:results-SFppv}.

The dynamical state of the clusters is investigated via calculation of
the virial ratio,
\begin{eqnarray}
Q_{i}=-\frac{T_{\star}}{\Omega} & = & \frac{3\sigma^{2}R}{2GM_{\star}},
\end{eqnarray}
where $T_{\star}$ and $\Omega$ are the total kinetic and gravitational 
potential energies of the stars, respectively. For a given radius $R$, 
$M_{\star}$ is the total enclosed stellar mass and $\sigma$ is the 1-D 
velocity dispersion of the enclosed particles.
Values of $Q_{i}<1$ indicate a bound cluster, while $Q_{i}=0.5$ represents
a state of virial equilibrium. 

We find virial ratios at $R_{1/2}$ of 0.34, 0.24, 0.29, and 0.19 for
the primary clusters from the simulations \texttt{B-1-1M-v00},
\texttt{d-1-v10}, \texttt{B-1-v10}, and \texttt{B-1-1M-v10},
respectively. These are all sub-virial, with collisions forming more
tightly bound clusters at this stage in their evolution. However, as
noted in \S\ref{S:stardyn}, due to $R_{1/2}$ approaching the grid
scale, gravitational forces are not well resolved and thus accurate
dynamical evolution of the cluster is limited.  When the virial ratio
is calculated at better-resolved scales of $R=0.4~{\rm pc}$, we find
increased values of 0.44, 0.38, 0.36, and 0.33 for the same clusters,
respectively.

\subsubsection{Angular Dispersion Parameter}
\label{subsec:ADP}

The \textit{angular dispersion parameter} (ADP), $\delta_{\rm
  ADP,N}(R)$ \citep{DaRio_ea_2014}, is a technique for quantifing the
degree of substructure of a stellar distribution, especially designed
for application to centrally-concentrated star clusters. It is similar
to the \textit{azimuthal asymmetry parameter} (AAP) developed by
\citep{Gutermuth_ea_2005}. In its simplest form, this technique
divides the distribution spatially into equal-area circular sectors
and compares the dispersion of the number counts contained within each
region. Further division using concentric annuli allows study of this
substructure as a function of radius. In order to account for a global
elongation or eccentricity of the cluster, best-fitted elliptical
annuli can be used. We will adopt this as our fiducial method. To
obtain the best-fit ellipse shape and orientation, a linear fit to the
stars projected in the central 0.2~pc of the cluster is used to set
the position angle, $\theta_{e}$ of the semi-major axis, $a$. Then the
dispersion in position in directions parallel and perpendicular to
$\theta_{e}$ are calculated to derive the eccentricity. For results in
a given annulus of semi-major axis, $a$, we display them at a radius,
$R$, for which the circular area would be equal to that of the
ellipse.

For a given annulus divided into a total of $N$ equal sectors, each
$i$th sector contains $n_{i}$ stars. The ADP is defined as:
\begin{equation}
\delta_{\rm ADP,N}=\sqrt{\frac{1}{(N-1)\overline{n}} \sum\limits_{i=1}^{N}(n_i-\overline{n})^{2}} = \sqrt{\frac{\sigma^{2}}{\sigma^{2}_{\rm Poisson}}},
\end{equation}
where $\sigma$ is the standard deviation of the $n_{i}$ values,
$\overline{n}$ is the average of the number of stars per sector in the
given annulus, and $\sigma_{\rm Poisson}$ is the standard deviation
expected from a Poisson distribution. Thus, values of $\delta_{\rm
  ADP,N} \simeq 1$ indicate nearly random distributions of sources,
azimuthally.

ADP analysis was performed using 20 equally-spaced concentric annuli
out to a maximum radius of 0.4~pc with $N=6$ equally-divided sectors.
$\delta_{\rm ADP,6}$ is computed for twenty orientations of the sector
pattern at every $3^{\circ}$ angular rotation, and the final value is
averaged.  Both circular and elliptical annuli are used to calculate
$\delta_{\rm ADP,6}$, using the same previously determined cluster
center.

The star cluster formed in the non-colliding model, which has a
relatively low number of stars and thus larger Poisson errors, has
$\delta_{\rm ADP,6} \simeq$ 1.5 - 2.5 for circular annuli and $\simeq
1$ - 2 for elliptical annuli.  In both cases, $\delta_{\rm ADP,6}$
peaks near $R=0.15~{\rm pc}$.

The primary clusters from the colliding models have similar
morphologies, especially the density-regulated and
magnetically-regulated $m_{\rm \star,min}=10~M_{\odot}$ cases.
In the \texttt{B-1-1M-col} model, there exists a denser population of
lower mass stars and the location of the subcluster $R=0.4~{\rm pc}$
is at a slightly different position. This slight deviation may be
attributed to the earlier onset of star formation from lower density
gas.

The circular and elliptical $\delta_{\rm ADP,6}$ values are similar,
although again the latter is slightly smaller in size.  For the
clusters in the \texttt{d-1-col} and \texttt{B-1-col} simulations,
$\delta_{\rm ADP,6}\simeq2$-3, while the \texttt{B-1-1M-col} model has
overall higher values of $\delta_{\rm ADP,6}\simeq6$-7.  The radial
behavior of the clusters from the three colliding models is similar as
well, in that the cluster has moderate values of $\delta_{\rm ADP,6}$
out to $R\approx 0.2~{\rm pc}$. These decrease at the outskirts of the
defined cluster, then increase relatively sharply out to $R\approx
0.4~{\rm pc}$ upon the presence of the subcluster.

\citet{DaRio_ea_2014} carried out a similar ADP analysis of the Orion
Nebula Cluster (ONC). They found $\delta_{\rm ADP,N}$ with $N=4,6,9$
rises from below 1 in the very center of the ONC to reach fairly
constant values of about 1.5 to 2.5 from 0.1~pc to about
1.3~pc. Accounting for ellipticity
in the annuli brings these values of $\delta_{\rm ADP,N}$ down to
about 1 to 1.5, with some variations near 2. Da Rio et al. concluded
the projected spatial distribution of young, embedded stars in the ONC
is relatively smooth, which may be evidence for dynamical processing
if the cluster is older than a few orbit crossing times.

The primary cluster formed in the \texttt{B-1-1M-nocol} model returns
similar values of $\delta_{\rm ADP,6}$ as the ONC, while the $m_{\rm
  \star,min}=10~M_{\odot}$ colliding models return slightly higher
values.  The \texttt{B-1-1M-col} case has much higher $\delta_{\rm
  ADP,6}$, which may be attributed to its much larger number of stars
($\sim10$ times higher) resulting in smaller Poisson errors.  As
discussed in \S\ref{S:stardyn}, we caution that the simulation code is
not able to resolve small scale gravitational interactions between
stars, potentially affecting the outcome of this metric. Still, we
expect future application of this technique to better resolved
clusters, including models where final stellar concentrations are
reduced by including local feedback, will provide useful comparisons
with observed young, embedded clusters to help test different
formation scenarios.

\subsubsection{Minimum spanning tree}

Another method of studying the hierarchical structure of stellar
distributions is through the use of the \textit{minimal spanning tree}
(MST). The MST \citep[developed for astrophysical applications
  by][]{Barrow_ea_1985} is a technique borrowed from graph theory in
which all of the vertices of a connected, undirected graph are joined
such that the total weighting for the graph edges is minimized. In the
case of star clusters, the projected euclidean distances between the
individual stars acts as the edge weight.

To study the hierarchical structure of a collection of stars,
\citet{Cartwright_Whitworth_2004} introduced a dimensionless
parameter, $Q$, which can distinguish and quantify between smooth radial
clustering (i.e., more centrally concentrated) vs. multi-scale type
clustering (i.e., more substructure). Specifically,
\begin{equation}
Q=\frac{\overline{s}}{\overline{m}}.
\end{equation}
The numerator is the normalized correlation length
\begin{equation}
\overline{s}=\frac{\overline{d}}{R_{\rm cluster}},
\end{equation}
where $\overline{d}$ is the mean pairwise separation distance between
the stars and $R_{\rm cluster}$ is the overall cluster radius,
calculated as the distance from the mean position of all stars to the
farthest star.
The denominator is the normalized mean edge length
\begin{equation}
\overline{m}=\sum_{i=1}^{N_{\star}-1}\frac{e_{i}}{\sqrt{(N_{\star}A)(N_{\star}-1)}} ,
\end{equation}
where $N_{\star}-1$ is the total number of edges, $e_{i}$ is the
length of each edge, and $A=\pi R_{\rm cluster}^{2}$ is the cluster
area.

\begin{deluxetable}{lcccccr}
\tablecaption{Parameters of Observed Clusters\label{tab:obs_q}}
\tablewidth{0pt}
\tablehead{
\colhead{Cluster} & \colhead{ } & \colhead{$Q$} & \colhead{ } &  \colhead{$\overline{s}$} & \colhead{ } & \colhead{$\overline{m}$} \\
}
\startdata
Taurus           & & 0.47  & & 0.55 & & 0.26  \\
IC2391           & & 0.66  & & 0.74 & & 0.49  \\
Chameleon        & & 0.67  & & 0.63 & & 0.42  \\
$\rho$ Ophiuchus & & 0.85  & & 0.53 & & 0.45  \\
IC348            & & 0.98  & & 0.49 & & 0.48  \\
\enddata
\edit2{\tablerefs{\citet{Cartwright_Whitworth_2004}}}
\end{deluxetable}

The threshold of $Q_{0}=0.785$ determines a quantitative threshold of
either smooth radial clustering ($Q>Q_{0}$) or multi-scale clustering
($Q<Q_{0}$). Table~\ref{tab:obs_q} lists $Q$, $\overline{s}$, and
$\overline{m}$ for various observed clusters \citep[see,
  e.g.,][]{Cartwright_Whitworth_2004}.

\begin{figure}[htb]
\centering
\includegraphics[width=1\columnwidth]{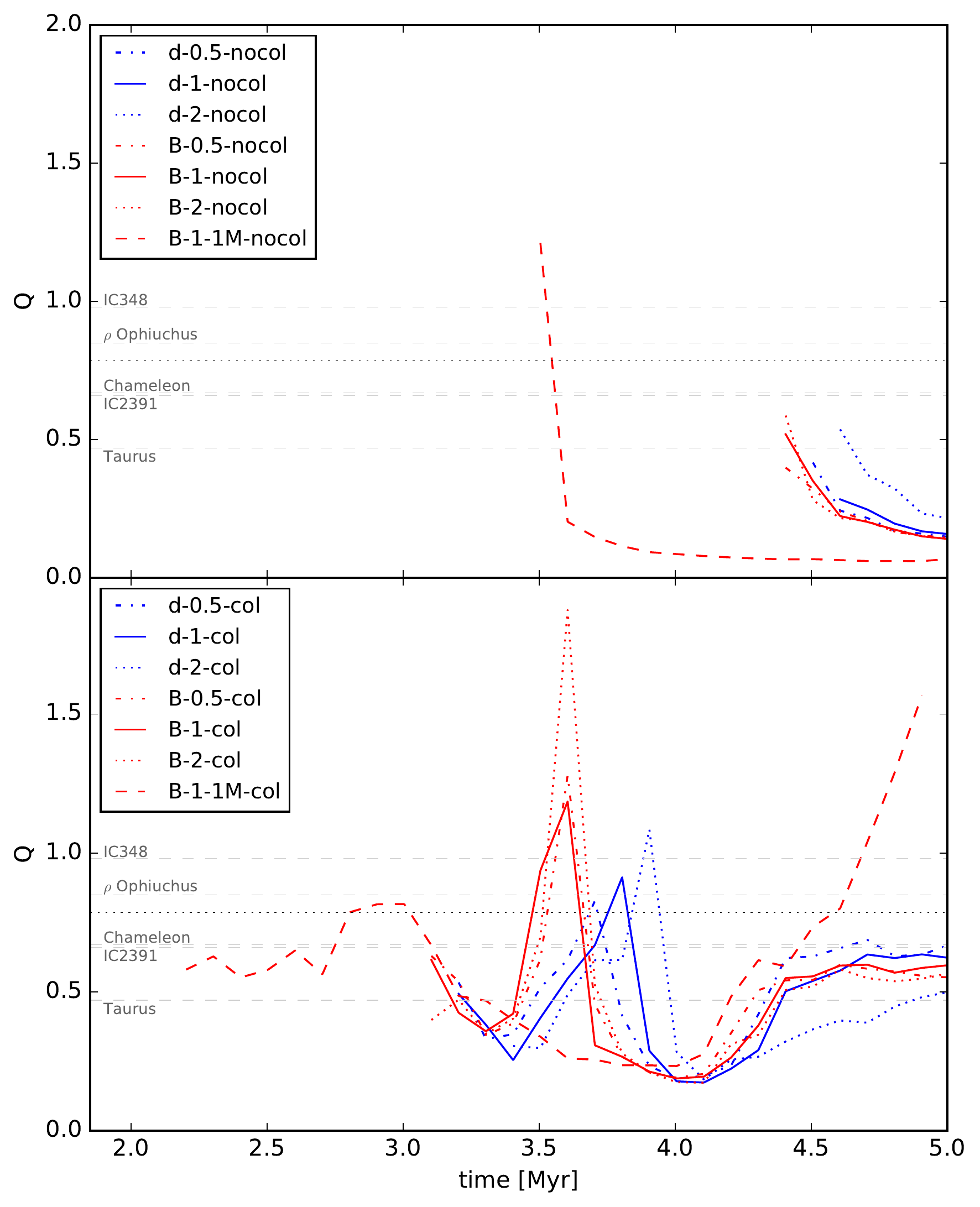}
\caption{
Minimum spanning tree $Q$ parameter vs. time. The evolution of $Q$ is
shown for each non-colliding (top) and colliding (bottom) model, as
denoted. The values of $Q$ are averaged over the three cardinal lines
of sight, $x$, $y$, and $z$. The dotted black line denotes the
threshold of $Q_{0}=0.785$, denoting smooth radial ($Q>Q_{0}$) vs multi-scale
($Q<Q_{0}$) clustering. The gray dashed lines show values of
$Q$-parameters of various observed star clusters from
Table~\ref{tab:obs_q}.
\label{fig:Q_vs_t}}
\end{figure}

We track the evolution of $Q$ throughout our simulations (see
Figure~\ref{fig:Q_vs_t}). In the $m_{\rm \star,min}=10\:M_{\odot}$
non-colliding cases, clustering begins with $Q\approx 0.5$, at the low
end of observed clusters (e.g., Taurus), and continues to decrease
monotonically. For \texttt{B-1-1M-nocol}, an initially much higher $Q$
is seen, but this quickly decreases to values even lower than the
other cases. This general behavior can be understood as the formation
of an initial cluster (more tightly concentrated in the $m_{\rm
  \star,min}=1\:M_{\odot}$ case). However, the overall stellar
distribution soon appears very dispersed as other independent clusters
form throughout the GMCs.

The colliding cases exhibit very different behavior. The
density-regulated cases begin with similar $Q$ parameter values and
initially decrease from 0.5 to 0.25. However, beginning near 3.4~Myr,
they experience a sharp increase in $Q$, reaching between 0.8 and 1.2,
corresponding with the high end of observed clusters (e.g., $\rho$
Ophiuchus and IC348). The higher-density cutoff models reach higher
maximum $Q$ values and peak at later times. After this peak, the $Q$
values drop to fairly multi-scale-type clustering, but then rise again
toward 0.5 to 0.7 in the final 0.5~Myr. The $m_{\rm
  \star,min}=10\:M_{\odot}$ magnetically-regulated cases have similar
qualitative behavior, with the $Q$ peak occurring earlier in time,
near 3.6~Myr and reaching very large values, surpassing centrally
clustered observations. However, $Q$ also drops down to $\sim 0.25$
but again equalizes to values near 0.6. These can be understood as the
initial formation of a moderately distributed star cluster which
quickly becomes very centrally dominated as a result of new star
formation in the compressed gas due to the collision that forms a
primary cluster. However, the colliding region soon produces other
clusters separated from the primary star cluster, thus decreasing
$Q$. Then, beginning from $t\approx 4.0$~Myr, the gas and stars
continue to coalesce, growing in size, number of stars, and central
concentration. The primary cluster grows and accumulates more of the
surrounding clusters, leading to the growth into a slightly multi-scale
distribution overall. The \texttt{B-1-1M-col} model begins near
$Q=0.6$ and experiences a lower peak near 0.8 before dropping off to
0.25. The final rise of $Q$ is concurrent with the other colliding
models, but instead of settling near 0.6, $Q$ continues to rise until
the end of the simulation $t=5.0$~Myr, reaching a very high central
clustering value of 1.5.

When comparing the results from our non-colliding vs. colliding
simulations, clusters formed by GMC collisions spend a much greater
fraction of the initial 5~Myr evolution with $Q$ parameters within the
range of observed clusters. While this result should not be
overinterpreted, as clusters produced in the non-colliding cases may
evolve into more centrally peaked distributions at beyond 5~Myr, a
much stronger clustering of stars naturally arises from colliding gas,
and this behavior is quantitatively realized in our simulations.

\subsection{Gas and Star Kinematics}
\label{sec:results-SFppv}

\begin{figure*}
  \centering
  \includegraphics[width=1\columnwidth]{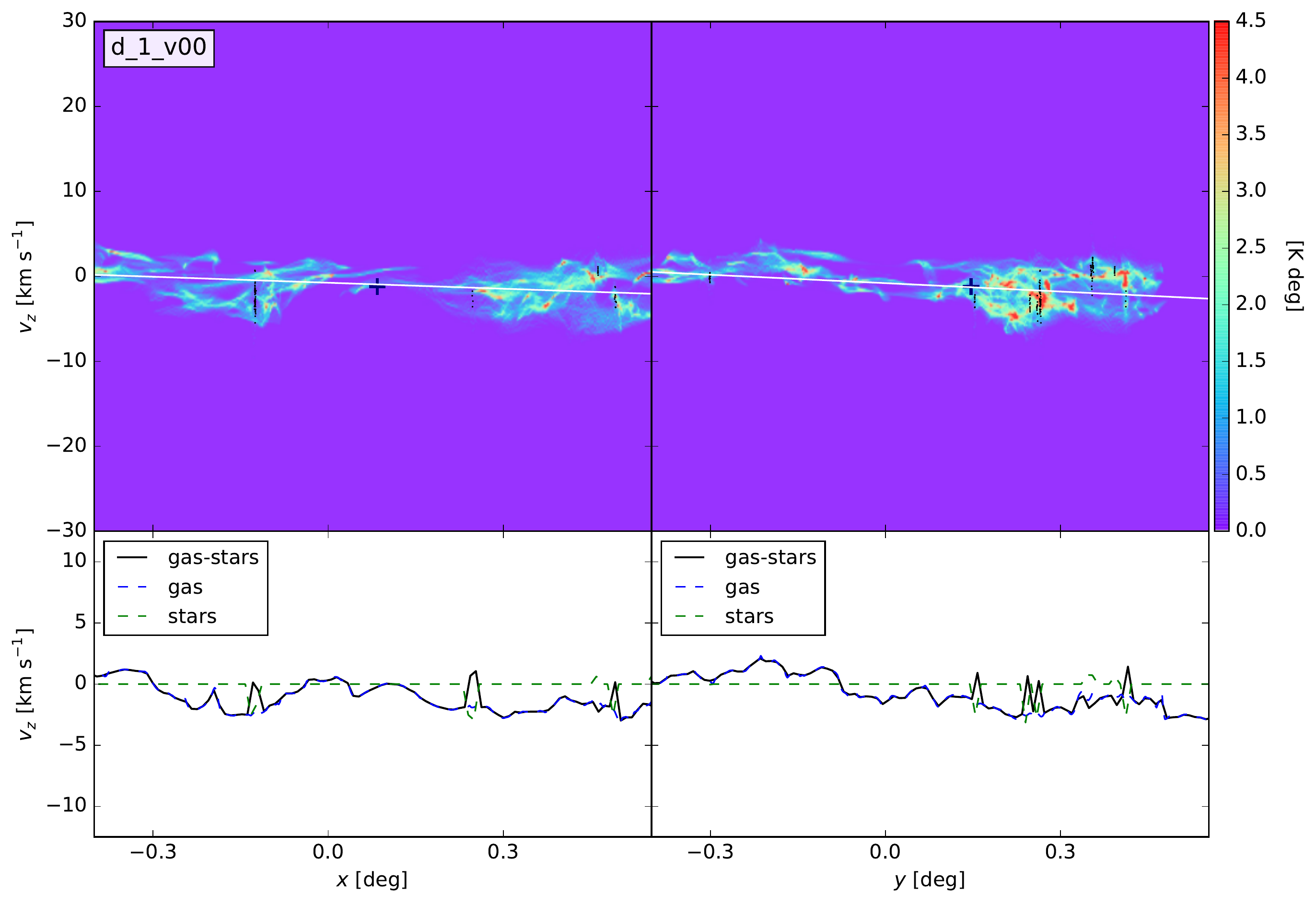}
  \includegraphics[width=1\columnwidth]{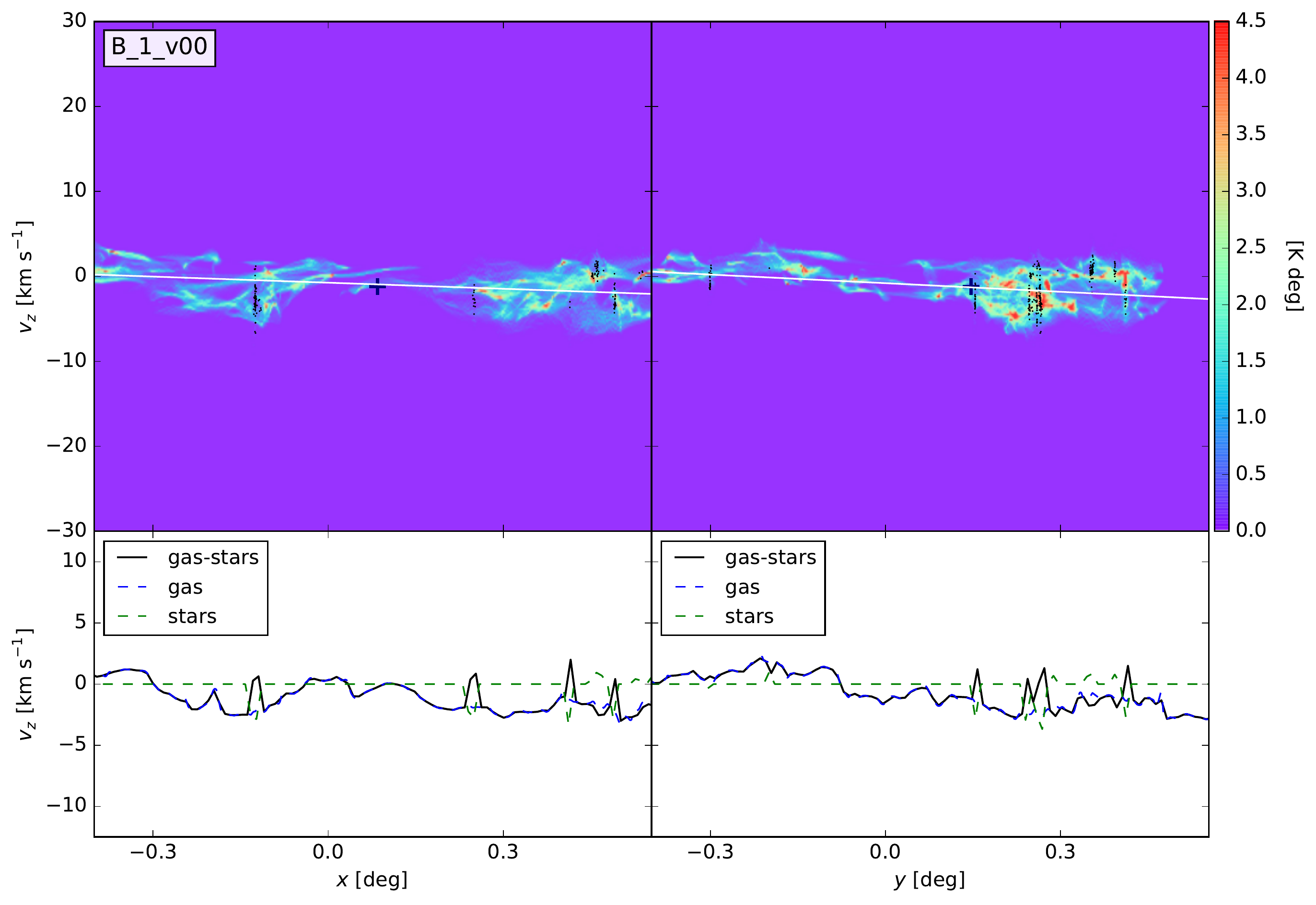}
  \includegraphics[width=1\columnwidth]{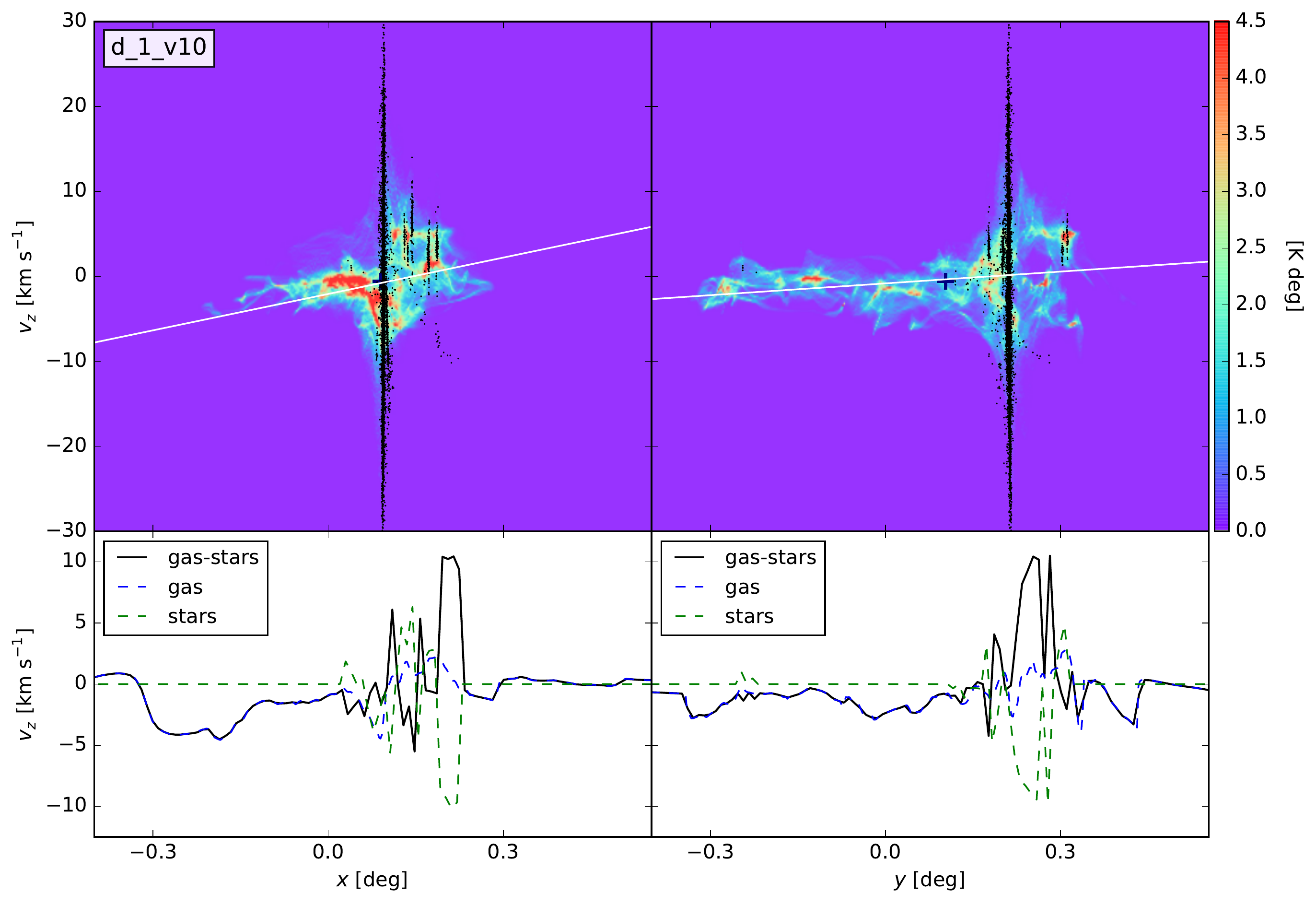}
  \includegraphics[width=1\columnwidth]{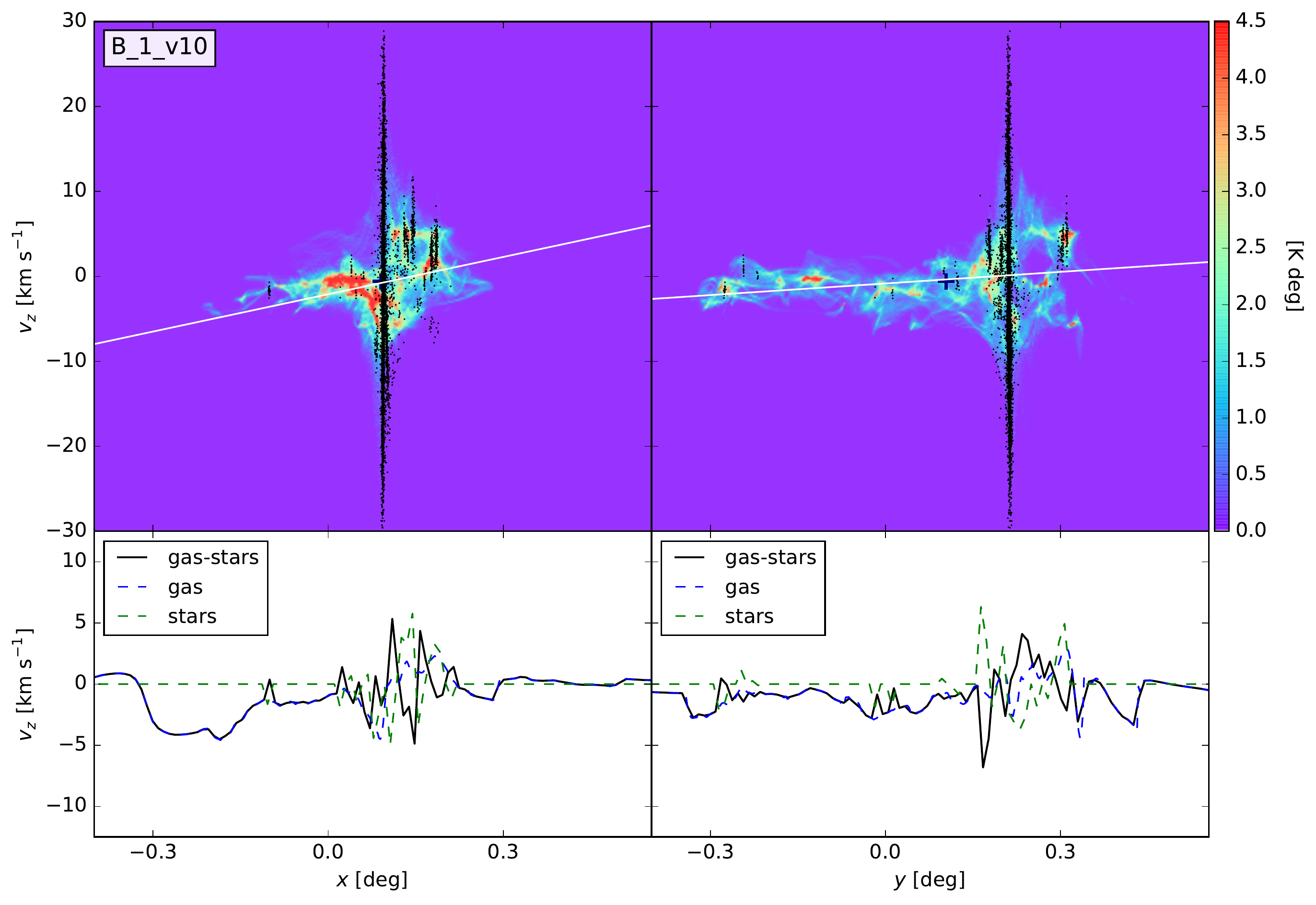}
\caption{
Position-velocity diagrams for selected non-colliding (top row) and
colliding (bottom row) simulations for the density-regulated (left
column) and magnetically-regulated (right column) star formation
routines. Each model is shown at $t=5.0\:{\rm Myr}$ along the $z$ line
of sight.  The colormap depicts synthetic $^{13}$CO($J$=1-0) line
intensities from the gas through velocity bins of $\Delta
v=0.212\:{\rm km\:s^{-1}}$.  The star particles are overplotted as
black points. The gray cross indicates the position of the center of
mass and the solid white line shows the intensity-weighted linear
velocity gradient ($dv_{\rm los}/ds$) across each cloud. Below each
respective position-velocity diagram are plots of the mean gas
velocity, mean star velocity, and their difference. Positional bins of
$0.5~{\rm pc}$ (i.e., $9.5\times 10^{-3}~{\rm deg}$ for an adopted
system distance of 3~kpc) are used.
\label{fig:SFppv}}
\end{figure*}

The relationship between the kinematics of young stars and their
surrounding gas has been studied in order to gain insight into the
formation and early evolution of young, embedded stellar
populations. For example, using data from the INfrared Spectra of
Young Nebulous Clusters (IN-SYNC) survey
\citep{Cottaar_ea_2014}, which achieves radial velocity accuracies of
about 0.3~$\rm km\:s^{-1}$, \citet{Foster_ea_2015} have studied the
kinematic properties of young stars in NGC~1333, while
\citet{Cottaar_ea_2015} have carried out a similar analysis of
IC~348. \citet{DaRio_ea_2017arXiv} \citep[see
  also][]{Hacar_ea_2016,Stutz_Gould_2016} analyzed similar data for
the ONC and its extended southern filament, including comparison to
gas tracers such as $\rm ^{13}$CO.  Our simulations allow a similar
investigation of the kinematic properties of both $^{13}$CO-defined
gas and the young stars under various star formation scenarios.

Our gas structures are defined using synthetic $^{13}$CO($J$=1-0)
emission, based on the same observational assumptions as Paper II
(i.e., GMCs are at a distance $d$=3~kpc, the optically thin limit
applies, and we bin with a spectral resolution of $0.212~{\rm
  km\:s^{-1}}$). Figure~\ref{fig:SFppv} shows position-velocity
diagrams for non-colliding and colliding cases for density and
magnetically-regulated star formation models 
\citep[such analysis methods have also been carried out in the 
simulations of][]{Duarte-Cabral_ea_2011,Dobbs_ea_2015,Butler_ea_2015,Haworth_ea_2015a}.
The mean gas velocity, mean stellar velocity, and difference in these
means as functions of position are also shown as profiles plotted
below their respective colormaps.  Mean values are taken using
positional bins of 0.5~pc (i.e., $9.5\times 10^{-3}~{\rm deg}$).

The non-colliding cases show widely dispersed gas over the positional
space, with clumpy morphology in $^{13}$CO($J$=1-0). The velocity
$v_{z}$ is fairly low, staying within $\pm 5~{\rm km\:s^{-1}}$. The
gas velocity gradient is relatively shallow, following the general
structure of the clouds. The gas and stellar kinematics in the
density-regulated and magnetically-regulated star formation models are
similar, with more star clusters present in the \texttt{B-1-v00}
model. The star clusters can be seen localized in positional space
with a small scatter in velocity space. Generally, the stars are
positioned in the vicinity of other high-intensity $^{13}$CO clumps.

The colliding cases show very different behavior in position-velocity
space. The gas is much more localized spatially in $x$, the direction
of the colliding flows. The structure is more concentrated in the
$y$-direction as well due to the higher central gravitational
potential formed. The average $^{13}$CO-weighted velocity gradient is
much steeper in $x$ for the colliding cases, but relatively similar in
magnitude to the non-colliding cases in the $y$ line of sight.  Larger
clumps with higher intensities of $^{13}$CO gas are seen in the
colliding cases, with a dense network of filamentary structures
present. Additionally, the gas velocity dispersion is much greater,
with portions reaching velocity dispersions of $\pm 20~{\rm
  km\:s^{-1}}$. Gas and stellar kinematic morphologies are also
similar among the different star formation models.  The central star
cluster is seen to have a very large velocity dispersion, with the
primary clusters in \texttt{d-1-col} and \texttt{B-1-col} models found
in \S~\ref{S:clust-props} to have $\sigma_{z,1/2}=9.94$ and $9.83~{\rm
  km\:s^{-1}}$, respectively.  Separate, smaller clusters can also be
seen with their own stellar populations near high-intensity clumps of
gas.

\begin{table*}
\caption{Gas and Star Kinematics}
\label{tab:SFkinematics}
\begin{center}
\begin{tabular}{lccccccc}
\hline\hline
Case        & LoS  & $\frac{dv_{\rm los}}{ds}$    & $\sigma_{\rm gas}$  & $\sigma_{*}$ & $\overline{v}_{\rm gas}$ & $\overline{v}_{*}$ & $\Delta \overline{v}$\\
            &      & ($\rm{km\:s^{-1}\:pc^{-1}}$) & ($\rm{km\:s^{-1}}$) & ($\rm{km\:s^{-1}}$) & ($\rm{km\:s^{-1}}$)  & ($\rm{km\:s^{-1}}$) & ($\rm{km\:s^{-1}}$) \\
\hline
d-1-v00    & $x$  & 0.096 & 1.810 &  1.700 &  1.024 &  0.705 &  0.319 \\
           & $y$  & 0.017 & 1.271 &  1.179 & -0.124 & -0.657 &  0.533 \\
           & $z$  & 0.077 & 2.107 &  1.882 & -1.189 & -1.358 &  0.169 \\
           & RMS  & 0.072 & 1.763 &  1.615 &  0.909 &  0.961 &  0.372 \\
B-1-v00    & $x$  & 0.096 & 1.810 &  1.777 &  1.022 &  0.554 &  0.467 \\
           & $y$  & 0.017 & 1.275 &  1.349 & -0.124 & -0.343 &  0.220 \\
           & $z$  & 0.079 & 2.110 &  2.055 & -1.188 & -1.450 &  0.261 \\
           & RMS  & 0.072 & 1.766 &  1.751 &  0.908 &  0.918 &  0.334 \\
d-1-v10    & $x$  & 0.194 & 4.087 &  8.301 & -0.305 &  0.810 & -1.115 \\
           & $y$  & 0.278 & 4.428 & 10.427 &  0.566 & -1.344 &  1.909 \\
           & $z$  & 0.287 & 4.093 & 10.683 & -0.583 & -1.226 &  0.643 \\
           & RMS  & 0.256 & 4.206 &  9.862 &  0.501 &  1.150 &  1.330 \\
B-1-v10    & $x$  & 0.194 & 4.059 &  8.307 & -0.324 &  0.881 & -1.206 \\
           & $y$  & 0.288 & 4.408 &  9.989 &  0.567 & -1.377 &  1.944 \\
           & $z$  & 0.293 & 4.104 & 10.170 & -0.590 & -1.171 &  0.582 \\
           & RMS  & 0.262 & 4.193 &  9.526 &  0.508 &  1.161 &  1.363 \\
\hline
\end{tabular}
\end{center}
\end{table*}

From the position-velocity information, we calculate velocity
gradients of the gas ($\frac{dv_{\rm los}}{ds}$), the velocity
dispersion of the gas ($\sigma_{\rm gas}$), the velocity dispersion of
the stars ($\sigma_{\rm *}$), the $^{13}$CO-weighted average
velocities of the gas ($\overline{v}_{\rm gas}$), the mass-weighted
average velocities of the stars ($\overline{v}_{*}$), and the velocity
offset between the two ($\Delta
\overline{v}$). Table~\ref{tab:SFkinematics} summarizes these
properties for the four models as viewed from $x$, $y$, and $z$ lines
of sight, as well as their RMS values.

We note that the average $\sigma_{*}$ values (9.86 and 9.53~${\rm
  km\:s^{-1}}$) in the whole domain are only slightly lower than those
of the primary cluster (9.94 and 9.83~${\rm km\:s^{-1}}$). In these
cases, the central cluster contains the majority of the stars and thus
dominates the overall distribution.

Within the non-colliding cases, both the gas and stellar kinematics
agree fairly closely between different star formation models. The
velocity gradients are larger in the $x$ and $y$ directions due to
asymmetries from the impact parameter. A RMS value of
$0.072~\rm{km\:s^{-1}\:pc^{-1}}$ is recorded. The dispersion of the
gas and stars are similar in both models, with $\sigma_{\rm gas,
  RMS}\approx 1.76~\rm{km\:s^{-1}}$ and $\sigma_{\rm *, RMS}\approx
1.68~\rm{km\:s^{-1}}$. The mean velocities of the gas and stars are
also similar, with $\overline{v}_{\rm gas, RMS}\approx
0.91~\rm{km\:s^{-1}}$ and $\overline{v}_{\rm *, RMS}\approx
0.94~\rm{km\:s^{-1}}$. The stellar velocities in the density-regulated
model have slightly lower dispersions, but higher mean
values. Overall, the velocity offset between the gas and the stars for
the non-colliding models is approximately $0.35~\rm{km\:s^{-1}}$.

For the colliding cases, the density-regulated and
magnetically-regulated star formation models exhibit similar kinematic
properties of the gas and stars. On average, $\frac{dv_{\rm
    los}}{ds}=0.26~\rm{km\:s^{-1}\:pc^{-1}}$, $\sigma_{\rm gas,
  RMS}=4.14~\rm{km\:s^{-1}}$, $\sigma_{\rm *,
  RMS}=9.70~\rm{km\:s^{-1}}$, $\overline{v}_{\rm gas,
  RMS}=0.50~\rm{km\:s^{-1}}$, $\overline{v}_{\rm *,
  RMS}=1.16~\rm{km\:s^{-1}}$, and $\Delta
\overline{v}=1.35~\rm{km\:s^{-1}}$. Relative to non-colliding clouds,
the collision induces a much larger velocity gradient ($\sim$3-4 times
greater), a larger velocity dispersion in the gas ($\sim$2 times
greater), and a much larger stellar velocity dispersion ($\sim$5 times
greater).

As functions of position, the mean velocities of gas, stars, and their
offsets, are compared. Offsets exist in both non-colliding and
colliding cases, becoming most apparent in close proximity to star
clusters.  For the non-colliding case, these differences are
relatively small, at a few ${\rm km\:s^{-1}}$. However, the colliding
case contains regions in which the offsets exceed $5~{\rm
  km\:s^{-1}}$.  Averaged over position space, colliding GMCs result
in velocity offsets a factor of $\sim$4 times higher than
non-colliding GMCs. This parameter may be an indicator for determining
the dynamical formation history of young star clusters, as results
from collisions show more disturbance kinematically.

We make a first, simple comparison with the results of the IN-SYNC
survey of the ONC and surroundings \citep{DaRio_ea_2017arXiv}. This
survey found offsets between gas and star velocities of approximately
$\Delta v_{r}\sim -0.5$~km/s, but up to $-1.0$ to $-1.5$~km/s in some
regions. The magnitude of such offsets are in general consistent with
those seen in both the non-colliding and colliding models, especially
considering variations associated with the particular line of sight.
Given that the observational data for Orion is just a single example
of a star-forming region, viewed on a particular sight line, it is
difficult to draw definitive conclusions about whether the gas and
star kinematics favor one scenario over another. Larger numbers of
star-forming regions need to be studied with similar methods. In
addition, other metrics, such as the comparison of low and high
density gas tracers \citep[e.g.,][]{Henshaw_ea_2013,Henshaw_ea_2014},
need to be examined, which on the simulation side requires extension 
of the astrochemical modeling to include species such as $\rm N_2H^+$.

\section{Discussion and Conclusions}
\label{sec:conclusion}

We have implemented two classes of star formation sub-grid routines
into the MHD code Enzo that we are using to study GMC collisions: a
density-regulated model based on a threshold density and a new
magnetically-regulated model based on a threshold mass-to-flux
ratio. Varying key parameters for each star formation routine, we
explored the large-scale morphology, properties of star-forming gas,
global star formation rates and efficiencies over time, spatial
clustering of the stars, and gas and stellar kinematics. For each
model, we investigated scenarios of non-colliding and colliding GMCs.

The non-colliding cases evolved in a relatively quiescent manner,
driven by the initial turbulence and interplay of self-gravity and
magnetic fields. Star clusters formed only in the very late stages of
the simulations, from overdense clumps located within filaments and
dispersed throughout the GMC complex. Generally, these clusters
contained hundreds of solar masses each and grew at a relatively slow
rate. Star clusters in the density-regulated star formation routines
were smaller and more isolated. The clusters formed in the
magnetically-regulated models exhibited slightly more elongated
morphologies. For this simulation set up, the level of star formation
activity appears to be regulated by the effective density threshold
that is used in each of these models, with the mass-to-flux criterion
not having a large influence.

During collisions between GMCs, stars formed earlier and in larger
clusters, from high-density gas produced in the primary filamentary
colliding region. 
While star formation rates level off by the completion of the 
simulations, extrapolation of future behavior is unclear. 
Nevertheless, by $t=5~{\rm Myr}$, individual clusters have grown
and merged to form one large, dominant cluster with a total stellar
mass of $5\times 10^{4}~M_{\odot}$. For this particular set-up, the
final overall level of star formation is relatively independent of all
the explored star formation sub-grid models. Star formation appears to
be limited by the ability of the collision to direct mass into high
density regions, which then eventually form stars with high overall
efficiency.

Just prior to star formation, both density and magnetically regulated
star formation result in fairly similar gas properties of parent
cells. However, colliding cases experience relatively wider ranges of
densities, temperatures, $\mu_{\rm cell}$, and velocity
magnitude. Higher mean values for density and temperature are found,
while gas is more magnetically subcritical and turbulent.

The primary star clusters formed in the various models were analyzed
and found to have much higher surface densities at their half-mass
scale than any observed cluster. We expect the future inclusion of
stellar feedback will reduce these surface densities. 
The angular dispersion parameter (ADP) analysis was carried out on the
primary clusters in the simulations. ADP values are generally greater
than those see in the ONC, which may indicate the ONC is dynamically
older than the simulated clusters.
The MST $Q$ parameter was also used to investigate the global spatial
distribution properties of the star, with non-colliding cases
resulting in overall highly multi-scaled clustering due to the
scattered formation of independent clusters. Colliding GMCs produce
clusters with $Q$ parameters that vary between those expected of
multi-scale and centrally clustered distributions.

Kinematically, our colliding GMC cases produce velocity gradients 3-4
times greater than those of the non-colliding cases. The velocity
dispersions also differ, with the gas in the colliding clouds having
approximately twice the velocity dispersion. Stellar velocity
dispersions in the simulations are dominated by the potentials of the
primary clusters that form, with this leading to much greater
dispersions in the colliding case. We find that the colliding cases
produce typically 4 times larger offsets between the mean gas and mean
star velocities compared to the non-colliding case.

Finally, we remind the reader of several important caveats. The young
stars do not inject feedback, especially protostellar outflow
feedback, into the surrounding gas. Internal star cluster dynamics are
not well followed because of gravitational softening at the grid scale
($\sim 0.1$~pc) and because the star particles lack realistic mass and
multiplicity distributions. Still the conditions that are simulated
here may provide boundary conditions for more detailed models that are
able to follow full $N$-body evolution of the clusters
\citep[e.g.,][]{Farias_2017arXiv}. Finally, in the context of GMC
collisions, a wide range of cloud (e.g., degree of initial
magnetization) and collision (e.g., velocities; impact parameters)
parameters remain to be explored with these models. These items will
be addressed in subsequent papers in this series.

\acknowledgments The authors would like to thank Nicola Da Rio and
Juan Pablo Farias for useful discussions.  Computations described in
this work were performed using the publicly-available \texttt{Enzo}
code (http://enzo-project.org), This research also made use of the
yt-project (http://yt-project.org), a toolkit for analyzing and
visualizing quantitative data \citep{Turk_ea_2011}. Volume renderings
were performed using \texttt{VisIt}
(https://wci.llnl.gov/simulation/computer-codes/visit). The authors
acknowledge University of Florida Research Computing (www.rc.ufl.edu)
for providing computational resources and support that have
contributed to the research results reported in this publication.

\software{Enzo (Bryan et al. 2014), Grackle (Bryan et al. 2014; Kim et al. 2014), PyPDR, yt (Turk et al. 2011), VisIt}

\clearpage

\end{document}